%% file: ooprompt.tex
\newcommand{\buildmode}{arxiv}

\ifnum\pdfstrcmp{\buildmode}{arxiv}=0
\newcommand{\mrevcolor}{black}
\else
\newcommand{\mrevcolor}{black}
\fi

\newcommand{\mrev}[1]
{{\color{\mrevcolor}{#1}}}

\ifnum\pdfstrcmp{\buildmode}{arxiv}=0
\documentclass[sigconf,nonacm]{acmart}
\else\ifnum\pdfstrcmp{\buildmode}{camera-ready}=0
\documentclass[sigconf]{acmart}
\else
\documentclass[manuscript,review,anonymous]{acmart}
\fi\fi

\usepackage{booktabs}   
\usepackage{tabularx}   
\usepackage{xcolor}
\usepackage{wrapfig} 
\usepackage{ragged2e}
\usepackage{array}
\usepackage{enumitem}
\usepackage{etoolbox}
\usepackage{placeins}
\usepackage{xac}
\usepackage{xspace}

\newlist{tabitemize}{itemize}{1}
\setlist[tabitemize]{label=\labelitemi, leftmargin=*, nosep, after=\vspace{-\baselineskip}, before=\vspace{-0.6\baselineskip}}

\newlist{tightitemize}{itemize}{1}
\setlist[tightitemize]{
    label=\textbullet, 
    leftmargin=*, 
    nosep,           
    after=\strut,    
    before=\vspace{-0.8\baselineskip} 
}
\newcolumntype{L}[1]{>{\RaggedRight\arraybackslash}p{#1}}
\newcolumntype{P}[1]{>{\RaggedRight\arraybackslash}p{#1}}
\newcolumntype{Z}{>{\RaggedRight\arraybackslash}X}

\setlist[tabitemize]{label=$\bullet$, leftmargin=*, nosep, after=\vspace{-\baselineskip}, before=\vspace{-0.8\baselineskip}}

\AtBeginDocument{%
  }

\ifnum\pdfstrcmp{\buildmode}{arxiv}=0
\setcopyright{none}
\acmDOI{}
\acmISBN{}
\acmConference{}{}{}
\else\ifnum\pdfstrcmp{\buildmode}{camera-ready}=0
\setcopyright{acmlicensed}
\copyrightyear{2026}
\acmYear{2026}
\acmDOI{XXXXXXX.XXXXXXX}
\acmConference[Conference acronym 'XX]{Make sure to enter the correct
  conference title from your rights confirmation email}{June 03--05,
  2026}{Woodstock, NY}
\acmISBN{978-1-4503-XXXX-X/2018/06}
\else
\setcopyright{none}
\acmDOI{}
\acmISBN{}
\acmConference{}{}{}
\fi\fi

\begin{document}

\ifnum\pdfstrcmp{\buildmode}{camera-ready}=0
\title{OOPrompt: Reifying Intents into Structured Artifacts for Modular and Iterative Prompting}
\else
\title{\mrev{OOPrompt: Reifying Intents into Structured Artifacts for Modular and Iterative Prompting}}
\fi

\ifnum\pdfstrcmp{\buildmode}{manuscript}=0
\author{Anonymous Authors}
\else
\author{Tengyou Xu}
\affiliation{%
  \institution{University of California, Los Angeles}
  \city{Los Angeles}
  \state{California}
  \country{USA}
}
\email{tengyoux@ucla.edu}

\author{Detao Ma}
\affiliation{%
  \institution{University of California, Los Angeles}
  \city{Los Angeles}
  \state{California}
  \country{USA}
}
\email{detaoma2002@ucla.edu}

\author{Xiang `Anthony' Chen}
\affiliation{%
  \institution{University of California, Los Angeles}
  \city{Los Angeles}
  \state{California}
  \country{USA}
}
\email{xac@ucla.edu}
\fi

\ifnum\pdfstrcmp{\buildmode}{camera-ready}=0
\renewcommand{\shortauthors}{Xu et al.}
\else\ifnum\pdfstrcmp{\buildmode}{arxiv}=0
\renewcommand{\shortauthors}{Xu et al.}
\else
\renewcommand{\shortauthors}{Anonymous Authors}
\fi\fi

\newcommand{\oop}{\textsf{OOPrompt}\xspace}

\begin{abstract}
  \input{00_abstract.tex}
\end{abstract}




\maketitle

\input{01_intro.tex}

\input{02_overview}

\input{03_related_work}

\input{04_formative_study}

\input{05_design_space}

\input{06_validation}

\input{07_discussion}
\bibliographystyle{ACM-Reference-Format}
\bibliography{references}


\end{document}

%% file: 00_abstract.tex



\rev{
The rise of large language models (LLMs) has given rise to a class of prompt-based interactive systems where users primarily express their input in natural language.
However, composing a prompt as a linear text string becomes unwieldy when capturing users' multifaceted intents.}
We present \textbf{Object-Oriented Prompting (\oop)}, an emergent interaction paradigm that enables users to create, edit, iterate, and reuse prompts as structured, manipulable artifacts, unifying and generalizing several existing point systems. We first outlined a design space from existing work and built an early prototype, which we deployed as a probe in a formative study with 20 participants. Their feedback informed an expanded \oop design space. 
We then developed the full \oop prototype and conducted a validation study to further understand \oop's added values and trade-offs.
\rev{We expect the \oop design space to provide theoretical and empirical guidance to the design and engineering of prompt-based, LLM-enabled interactive systems}.


%% file: 01_intro.tex
\section{Introduction}
Large language models (LLMs) promise a powerful way for people to interface with software by simply describing their intent in natural language,
\rev{which, in turn, fosters an emergent class of interactive systems where the primary input is prompt-based.}

However, composing an effective prompt is challenging.
Users often think of prompts as a linear string of tokens, but their intent may encompass multiple aspects, such as goals, constraints, roles, personas, context, and expected output format~\cite{aiinstruments, sensecape, kim2023cellsgeneratorsandlenses, whyjohnnycantprompt}. 
Current prompt-based interactive systems often default to a one-shot, unstructured input, which makes it difficult for users to recognize, decompose, or iterate on the various key aspects of their intent~\cite{mao2025promptstemplatessystematicprompt, wang2024langgptrethinkingstructuredreusable, schulhoff2025promptreportsystematicsurvey, drosos2024dynamic, ma2024you}. 
\rev{As a result, many users' prompts only partially and sub-optimally represent their intent, thus limiting the quality of LLM's response and the overall experience with the interactive system 
\cite{ma2025should, shen2025interactionaug, joshi2024coprompter, khot2023decomposedpromptingmodularapproach, chainingllmvisual, reynolds2021prompt, the_design_of_everyday_things}.} 

Recently, we have witnessed a paradigm shift from {linear, one-shot prompts} toward more {structured, manipulable artifacts that externalize user intent}, as demonstrated in several point systems.
\rev{
Examples include 
enabling iterative prompting by directly manipulating interface objects \cite{aiinstruments, masson2024directgpt} and
employing stateful UI controls (\eg radio buttons, checkboxes, and text fields) for deterministic prompt editing~\cite{drosos2024dynamic, petridis2023promptinfuser},
as well as breaking down prompts into categorical dimensions to populate an explorable design space~\cite{suh2024luminate, chen2025genui}.
}
Perhaps the most relevant work is \textit{LangGPT}, a conceptual framework that treats prompts as a programming language. It suggests classifying user intents into high-level ideas (``modules'') and specific constraints (``elements''); prompts can then be structured as templates, with intents organized as instance variables in programming objects.

Together all this prior work indicates an emerging interaction paradigm of \textbf{Object-Oriented Prompting (\oop)}, where the intents involved in the prompt, as well as the logical relationships between them, are reified into interactive objects that encapsulate adjustable properties and support for iteration.
Although prior work has implemented specific facets of this paradigm, they remain limited as \textit{point systems} tailored to a single medium or workflow. 
Much as Shneiderman’s seminal essay unified disparate graphical interfaces under the ``direct manipulation'' banner \cite{directmanipulation}, this paper aims to provide the missing formalization of \oop.

\begin{figure*}[!tb]
    \centering
    \includegraphics[width=0.9\textwidth]{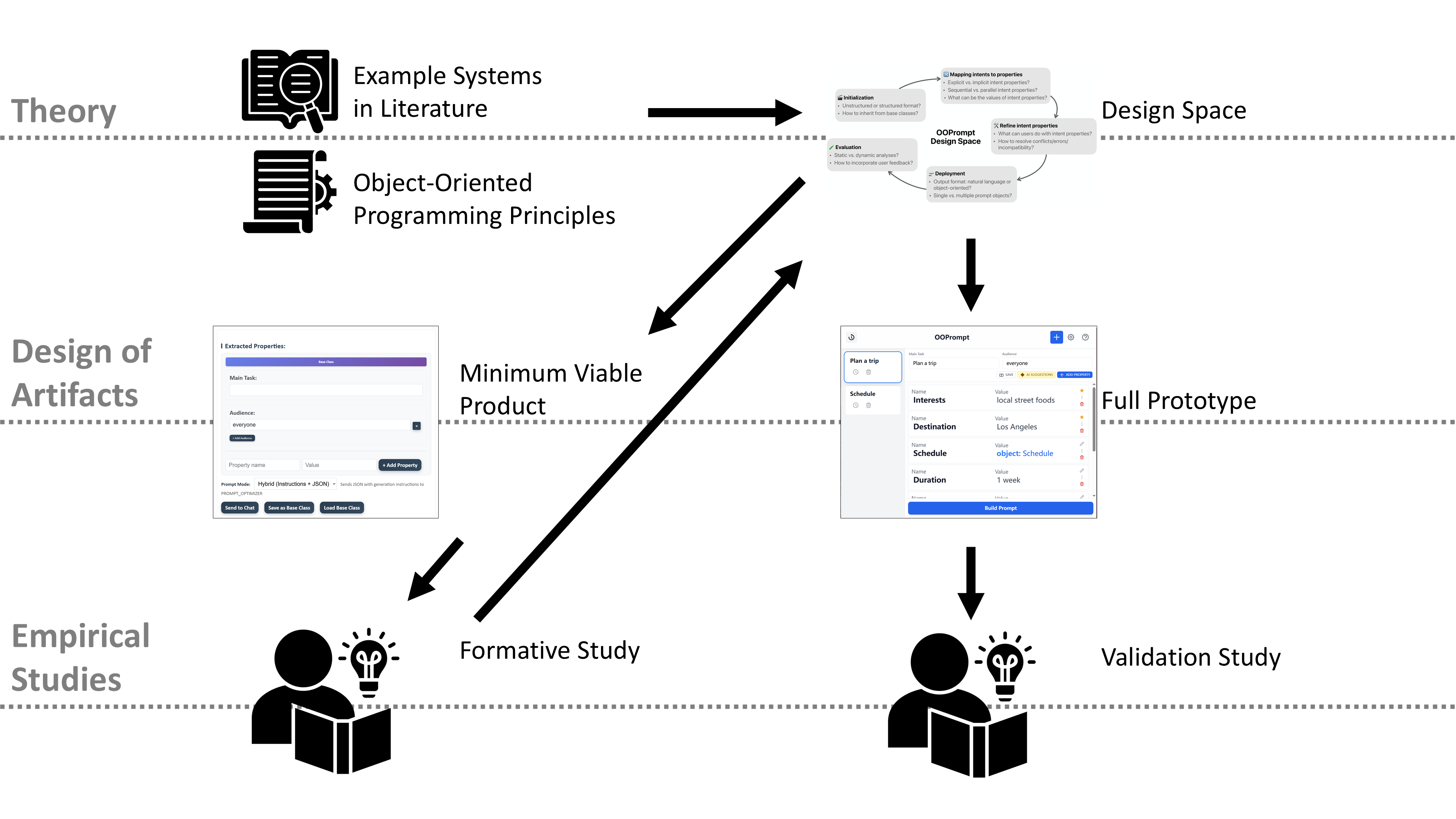}
    \caption{\textbf{Overview of research process} for developing Object-Oriented Prompting (\oop), progressing through six stages following the methodology in \cite{10.1145/263552.263612}: initial conceptualization based on existing systems and principles of object-oriented programming (OOP), minimum viable product (MVP) development, formative study using MVP as a probe, design space synthesized from formative study findings, development of a full \oop prototype, and its validation study.}
    \label{fig:flow_chart}
\end{figure*}

Fig.~\ref{fig:flow_chart} is an overview of our key research activities following the methodology in \cite{10.1145/263552.263612}. 
Synthesizing insights from recent \oop-related work, \rev{we first built an initial design space for \oop, based on which we then developed a minimum viable product (MVP)} that helps users compose \oop to send to conversational language models (\eg ChatGPT~\cite{achiam2023gpt}, Gemini~\cite{team2023gemini}). 
\rev{To construct the full \oop design space,} we used this MVP as a probe in a formative study~\cite{hutchinson2003technology} with $20$ participants who used the MVP to perform representative prompting-based tasks (\eg writing a short story, planning a trip). 
We then thematically analyzed participants' feedback to derive a comprehensive design space of \oop. 
Structured around the lifecycle of creating, reusing, editing, deploying, and evaluating an \oop, the design space outlines key design issues to address at each step when engineering a prompt-based interactive system. 


%
\rev{Following the \oop design space, we continued with further development of a full prototype and conducted a follow-up study with $8$ returning participants to validate \oop's added values and trade-offs.} 
\rev{Participants went through three in-depth task scenarios using the \oop prototype: reporting technical results to diverse audiences, planning a group trip event, and debating with varying opinions}. 
Each scenario included multiple sub-tasks, varying requirements, and/or different target audiences, followed by a brief interview about their experiences using \oop. 
\rev{Findings indicate that, by turning prompts into modular, editable properties, \oop makes planning and iterative writing faster and clearer, and better supports ambiguous, exploratory tasks, than traditional prompting.
Meanwhile, \oop struggles with opinion-forming reasoning and introduces extra steps that add latency, disrupt chat continuity, and require more automation to streamline integration into human-LLM interaction.
}

\rev{The main contribution of this research is the \oop design space, which is expected to \textbf{provide theoretical and empirical guidance for the design and engineering of prompt-based, LLM-enabled interactive systems}.
Rather than defaulting to an oversimplified prompt box, a developer can adopt the \oop design space as a specification for implementing a dialog for users to more fully and thoughtfully express their intent.}




%% file: 02_overview.tex
\section{Overview of \oop and a Comparison with OOP}
Formally, we define \oop as a paradigm for interacting with LLM-enabled systems, where the intents involved in a prompt, as well as the logical relationships between them, are reified into interactive objects with adjustable properties and support for iterative refinement.
\mrev{\oop leverages \textbf{Object-Oriented Programming (OOP)} as a conceptual metaphor to provide a structured framework for users to address abstract, complex intents. Rather than a literal transplantation of the ``object-oriented (OO)'' framing from programming to prompting (which may be neither feasible nor necessary), \oop incorporates the foundational properties of OOP as its core design rationales.} Table~\ref{tab:oop-vs-ooprompt} illustrates how \oop relates to the four fundamental pillars of OOP (i.e., Encapsulation, Abstraction, Inheritance, and Polymorphism)~\cite{booch2007oop}.

Specifically, \oop treats a prompt as a root-level object (\ie, a {prompt object}) composed of a collection of discrete, addressable units (i.e., {properties}). In this way, the prompt is no longer a monolithic block of text, but is decomposed into a set of intents. Each {property} functions as a container that encapsulates a specific aspect of user intent along with its associated metadata (\eg, emphasis weights, stylistic constraints, and few-shot examples), ensuring that the internal logic of one intent does not functionally interfere with another.
Further, \oop supports recursive decomposition of a complex task, allowing the value of a property to be either a primitive text string or a reference to another child {prompt object}. This recursive relationship enables the system to maintain structural consistency across varying levels of task complexity and helps users manage sophisticated hierarchical intents through a unified, modular interface.

\begin{table*}[t]
\centering
\small
\color{\revcolor} 
\caption{\rev{\textbf{Conceptual mapping between OOP and \oop}, based on the four fundamental pillars of OOP \cite{booch2007oop}}}
{\setlength{\tabcolsep}{4pt}
\begin{tabularx}{\textwidth}{@{} L{0.14\textwidth} L{0.30\textwidth} Z @{} }
\toprule
\textbf{Principle} & \textbf{Object-Oriented Programming (OOP)} & \textbf{Object-Oriented Prompting (\oop)} \\ \midrule
\textbf{Abstraction} & 
Structuring complex systems by modeling classes based on essential properties and behaviors. & 
Structuring the prompting of subtle, complex intent by decomposing it into the specification, editing, and suggestion of properties. \\
\addlinespace
\textbf{Encapsulation} & Bundling data and methods that operate on that data within a single unit (class), and restricting direct access to some of an object's components. & 
Bundling aspects of a prompt's underlying intent within a single interactive dialog with little restriction to directly access the prompt object's components. \\
\addlinespace
\textbf{Inheritance} & 
Allowing new classes to be based on existing classes, inheriting their attributes and methods. &
Allowing new prompt objects to be based on existing templates, inheriting their intent properties and built-in supports such as editing and suggestions. \\
\addlinespace
\textbf{Polymorphism} & 
Enabling objects of different types to be treated as objects of a common base type. & 
Enabling different types of intents to be conveyed via a single prompt object. \\
\bottomrule
\end{tabularx}
}
\label{tab:oop-vs-ooprompt}
\end{table*}

%% file: 03_related_work.tex
\section{Related Work}
\label{sec:related_work}



Recent work in the field of Human-Computer Interaction (HCI) has been exploring innovative interactive paradigms and user interface designs that enhance prompting. Studies point out the gap between human articulation and AI/LLM cognition, usually due to natural language's (NL) nature of ambiguity \cite{shen2025interactionaug, kim2023cellsgeneratorsandlenses}, redundancy \cite{shen2025interactionaug}, and fragility \cite{ wang2024langgptrethinkingstructuredreusable, jiang2020can}.  
NL addresses niche cases related to open-ended or nuanced requirements \cite{Liu2024structuredoutput}. 
    However, communication through NL prompts alone often cannot guide LLMs to meet users' expectations.
For example, NL prompts sometimes include overgeneralized instructions, causing LLM-generated content to be inaccurate or underspecified \cite{whyjohnnycantprompt, mishra2022reframing, subramonyam2024bridging}.
Moreover, freeform NL prompts often fall short in helping users explore ways to express complex intents \cite{suh2024luminate}, and they can be inefficient for iterative prompt modification \cite{wang2024langgptrethinkingstructuredreusable, mishra2025promptaidpromptexplorationperturbation, whyjohnnycantprompt, chainforge}.




To address these challenges, we review related work into two following categories:
\begin{itemize}
    \item \textbf{conceptual prompt engineering principles}: strategies and rules for composing effective prompts.
    \item \textbf{interactive interface design examples}: exemplar systems that help end users express abstract ideas through prompts.
\end{itemize}

\subsection{Conceptual Prompt Engineering Principles} 
Past studies have revealed meaningful principles on constructing prompts for effective communication with LLMs. 
    
Among the many techniques developed as LLMs have rapidly advanced, ``decomposition'' has become a dominant approach for designing more instructive and manipulable prompts. One classic study has shown that decomposition can significantly improve LLM performance by breaking complex tasks or reasoning into multiple serial steps, often known as chain-of-thought (CoT) prompting \cite{wei2023chainofthoughtpromptingelicitsreasoning,khot2023decomposedpromptingmodularapproach,aichains}. Systems that leverage this concept demonstrate strong potential for improving the efficiency and effectiveness of communication with LLMs. For example, \textit{DECOMP} by Khot et al. further develops the CoT concept and establishes an effective few-shot paradigm for solving complex tasks through hierarchical, recursive decomposition~\cite{khot2023decomposedpromptingmodularapproach}. In addition to chronological decomposition, this motivation has also been extended to more conceptual and abstract exploration.
Recent studies have been progressing to not only decomposition but also reconstruction. \rev{Some studies form structured prompts by summarizing from diverse real-world prompt examples and designing general prompt structures, as an effective way of reorganizing the extracted components after decomposing the initial prompt~\cite{mao2025promptstemplatessystematicprompt, chainforge, reynolds2021prompt, xu2024jamplate}.}
Perhaps the closest work is \textit{LangGPT}, which first associates prompts with OOP to leverage the fact that programming languages are designed for efficient communication with machines. Specifically, it proposes a ``dual-layer prompt design framework'' that conceptualizes prompts as software projects, with modules and elements analogous to classes and methods {to hierarchically address high-level intents and specific instructions}~\cite{wang2024langgptrethinkingstructuredreusable}. 
However, \textit{LangGPT} mainly focuses on the conceptual framework and validating it on existing LLMs; meanwhile, it remains unclear how user interfaces can appropriately support users to apply this idea to compose prompts as objects.
Our work builds on the same OOP foundation and fills in this gap by formalizing the \oop interaction paradigm.

\subsection{Interactive Interface Design Examples}
While studies in AI and prompt engineering provide many useful strategies for composing effective prompts, supported by both theory and experiments, a critical gap remains between the expertise required to apply these strategies and the accessibility needed for general end users to benefit from them. Simply forcing users to articulate their thoughts in a machine-preferred format is often neither efficient nor practical. The design of systems with proper interactions, therefore, is crucial to resolve this problem. 

Recently, the dominant perspective has focused on guiding users to build structured, reusable, and interactive artifacts from raw prompt text. Comprehensive studies have shown that users often prefer graphical interfaces for specifying low-level, quantifiable constraints, especially in tasks such as information seeking~\cite{Liu2024structuredoutput, graphologue}. We observe many valuable system paradigms following this approach. \textit{PromptChainer} provided an early paradigm of visual programming interfaces for building chains of LLM prompts and rendering them as node-edge diagrams, helping users decompose complex tasks with improved transparency and controllability~\cite{chainingllmvisual}. Similar designs have been further developed by other studies for better step-by-step planning, combinatorial querying, and organization of hierarchical logical relationships~\cite{chainforge, cai2024lowcodellmgraphicaluser, sensecape, joshi2024coprompter, wu2023autogenenablingnextgenllm}. Focusing on helping users restructure and process rough prompts, recent research extends interaction design toward element-level and dimension-level interaction, modification, and exploration, including possibilities that may not be explicit in the initial user prompt~\cite{han2025poet, suh2024luminate}. For example, Shen et al. provide a systematic framework, ``4W1H'' (``Why, When, Who, What, and How''), to analyze paradigms for designing interactions and instructions for generative AI~\cite{shen2025interactionaug}. \textit{PromptAid} and \textit{PromptPilot} import AI-powered modules to offer automated or semi-automated suggestions for vague or inaccurate wording~\cite{mishra2025promptaidpromptexplorationperturbation,promptpilot}. \textit{Cells, Generators, and Lenses} presents a typical approach that identifies key elements (``cells'') for LLMs, then reassembles them to create multiple generators for parallel prototyping~\cite{kim2023cellsgeneratorsandlenses}. Similarly, \textit{Luminate} highlights convergent-divergent thinking, helping users expand the idea space when building prompts~\cite{suh2024luminate}. Looking more closely at workflows, we also note studies that contribute effective interaction techniques within their systems. \textit{Dynamic Prompt Middleware} uses graphical elements such as buttons and checkboxes to replace direct text input~\cite{drosos2024dynamic}. \rev{Similarly, \textit{DirectGPT} enables direct manipulation to improve interaction with LLMs by replacing verbose prompts with physical actions on interactive interface elements~\cite{masson2024directgpt}.} \textit{AI-Instruments} notably transforms AI capabilities into an embedded toolkit and embodies prompts as graphical interface objects, enabling text-independent manipulations such as ``lenses'' and ``brushes''~\cite{aiinstruments}. This work demonstrates effectiveness in prompting abstract requirements that are difficult to describe clearly with words, echoing earlier object-oriented interaction ideas such as Attribute Objects in \textit{Object-Oriented Drawing}~\cite{oodrawing}. 
Collectively, these works contribute meaningfully to the development of interactive prompting systems, yet most target one specific task type or focus on a small set of augmentation techniques. Our \oop work aims to construct a generalizable design space that addresses a wide range of design issues and task scenarios, integrating techniques to optimize prompts across different stages of the prompt lifecycle.


%% file: 04_formative_study.tex
\section{Formative Study}
\label{sec:formative_study}
\rev{
In this study, we formulated the \oop design space by first synthesizing state-of-the-art \oop-like systems, which provided an initial version that we continued to build on by incorporating empirical insights from participants interacting with a minimum viable product (MVP) implementation of \oop.

Specifically, we aimed to probe the following research questions:

\begin{itemize}

\item[\textbf{RQ1}] \rev{How would users perceive and react to the high-level concept of \oop?}
\item[\textbf{RQ2}] How to develop and expand a comprehensive design space for \oop?
\end{itemize}

}


\subsection{Initial Version of the Design Space Based on Literature Analysis}

\begin{figure}[tbp]
    \centering
    \includegraphics[width=0.95\linewidth]{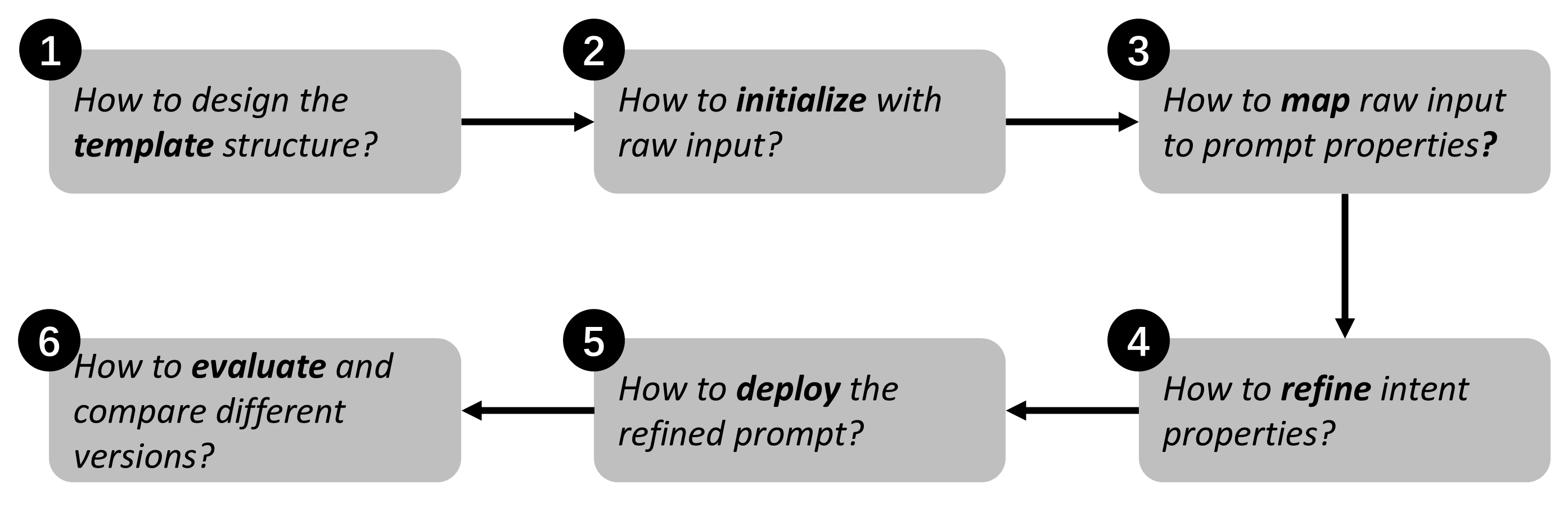}
    \caption{\textbf{The initial design space of OOPrompt}, outlining the key stages throughout the lifecycle of \oop artifacts.
    }
     \label{fig:initial_design_space}
\end{figure}

\rev{We began with analyzing a selected set of papers \cite{chainingllmvisual, shen2025interactionaug, mishra2025promptaidpromptexplorationperturbation, kim2023cellsgeneratorsandlenses, suh2024luminate, drosos2024dynamic, aiinstruments, oodrawing}, which presented example systems and workflows closest to the core concept of \oop.}
\rev{Our analysis followed a bottom-up process. First, we identified each paper's interaction design or technique related to \oop.
Then, by synthesizing these individual features, we constructed an initial design space that conceptualizes the \oop lifecycle across six common stages, as illustrated in \textbf{Fig.~\ref{fig:initial_design_space}}: (1) \textbf{Templating}, concerning the construction of structured ``base classes'' that serve as templates to pre-fill a prompt; (2) \textbf{Initialization}, where users instantiate an \oop object via free-form text input; (3) \textbf{Mapping}, which reifies the underlying user intent into discrete, structured properties; (4) \textbf{Refinement}, involving the iterative editing and optimization of these intent properties; (5) \textbf{Deployment}, where the finalized prompt is formatted and sent to LLM; and (6) \textbf{Evaluation}, which provides users with feedback to facilitate further prompt optimization.}



\subsection{Minimum Viable Product}


\begin{figure*}[!tb]
    \centering
    \includegraphics[width=1.0\textwidth]{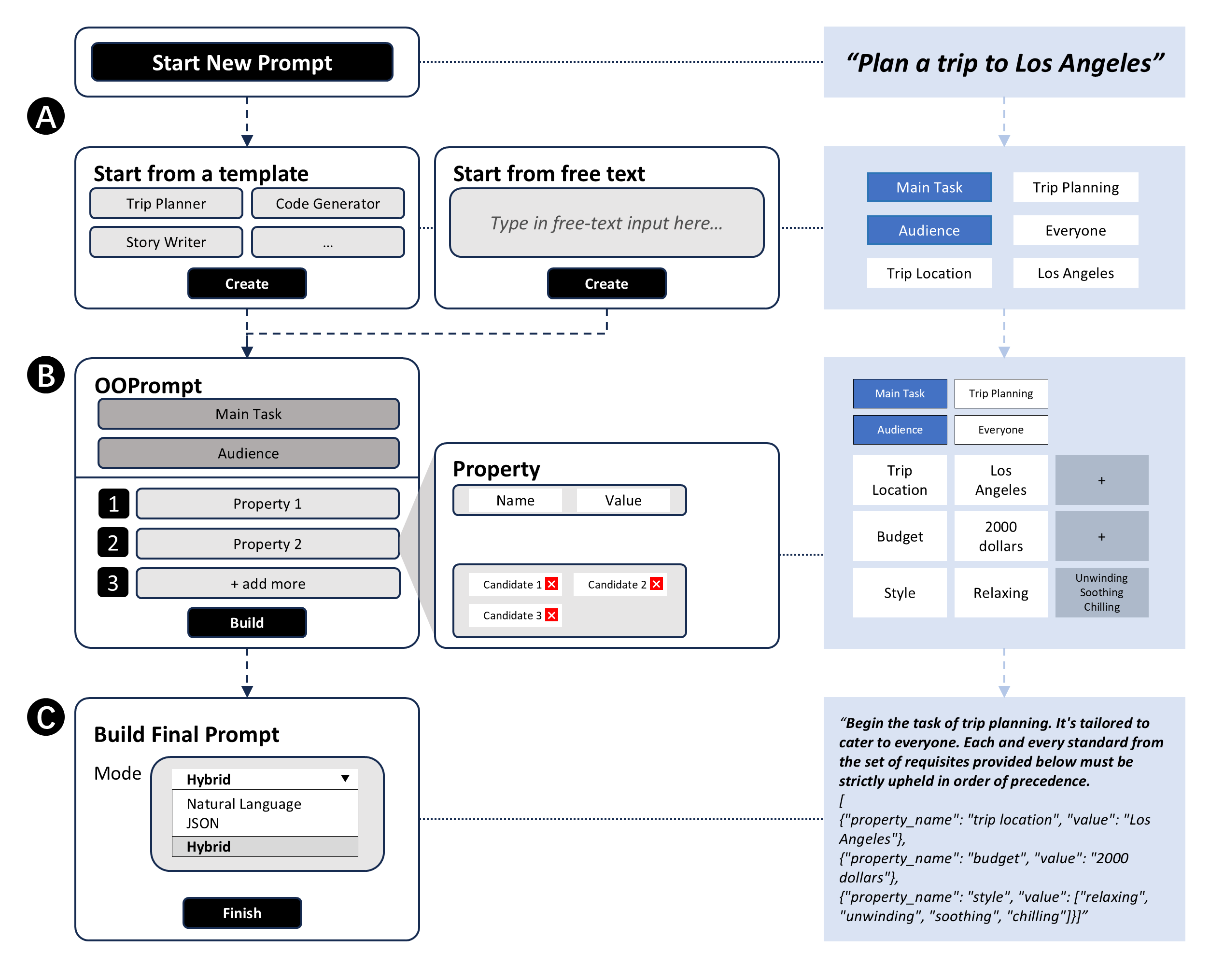}
    \caption{\rev{\textbf{The \oop MVP workflow with a simple walkthrough example about trip planning.} The diagram illustrates the end-to-end process of how user can interact with the interface, from initializing a prompt via templates or free-text input (A), through structuring and refining intent properties (B), to generating a final deployable prompt in natural language, JSON, or hybrid form (C). The example on the right demonstrates how a high-level request (\eg trip planning) is progressively decomposed into structured, editable properties and compiled into a final prompt. }}
    \label{fig:mvp_flow}
\end{figure*}

Fig.~\ref{fig:mvp_flow} shows the interface of the MVP, illustrating which functionalities the MVP supports and how users may interact with it. The basic workflow covered three core modules: \textbf{creating}, \textbf{editing}, and \textbf{deploying} an \oop to LLM:

\begin{itemize}    
    \item Create
    \begin{itemize}
        \item Unstructured: input free text and let the system extract key information (\ie, intent properties).
        \item Structured: select and load an existing base class as a ``template'' to fill out.
    \end{itemize}   
    \item Edit
    \begin{itemize}
        \item Add new properties by name-value pairs.
        \item Delete existing properties.
        \item Edit the name and/or values of existing properties.
        \item Generating different ways of describing existing properties (``candidates'').
        \item Manually rank the properties to indicate priority.
    \end{itemize}
    \item Deploy
    \begin{itemize}
        \item Generate a final NL prompt based on all properties and send it to LLM.
        \item Send the \oop in JSON format to LLM. 
        \item Generate a prompt in a hybrid format that combines NL and JSON.
    \end{itemize}   
\end{itemize}


\begin{table*}[t]
\centering
\caption{\textbf{Backgrounds and typical usage of AI/LLMs of the participants in formative study.} Participant IDs marked with * indicate users with an engineering background 
}
\resizebox{\textwidth}{!}{
\begin{tabular}{|c|l|l|p{7.5cm}|}
\hline
\textbf{ID} & \textbf{Background} & \textbf{LLM Use Frequency} & \textbf{Common Use Cases} \\
\hline
1*  & Electrical Engineering  & Daily & Coding, developing \\
2*  & Electrical Engineering & Daily & Coding, developing \\
3  & Social Science         &  4-6 times per week      & Writing assistance \\
4  & Jazz Music            &  Rarely & N/A \\
5  & Marketing             &  4-6 times per week      & Slides modification, copywriting \\
6*  & Computer Science    &  Daily & Coding, email modification \\
7  & Biomedical            &  4-6 times per week     & General information searching \\
8*  & Data Science         & Daily & Coding, information searching, email writing \\
9*  & Computer Science     & Daily & Coding, information searching \\
10 & Data Analysis       &  2-3 times per week    & Daily searching \\
11 & Business Analysis    & Weekly       & Chatting \\
12 & Statistics            & Daily   & Information Searching \\
13* & Mechanical Engineering & 4-6 times per week & Information searching, text generation \\
14 & Statistics             &  2-3 times per week    & Learning, information searching, writing \\
15* & Method of Computation  &  Daily & Coding, code explanation, information searching \\
16* & Computer Science       &  4-6 times per week & Coding, developing \\
17* & Computer Science & 4-5 times per week & coding, developing \\
18* & Electrical Engineering & 4-5 times per week & Coding, information searching \\
19* & Electrical Engineering & Daily & Coding, developing \\
20* & Computer Science & Daily & Coding, information searching \\
\hline
\end{tabular}
}
\label{tab:participants}
\end{table*}

\subsection{Participants}



We recruited 20 participants from 11 universities/companies (6 in the United States, 2 in Canada, 1 in Japan, 1 in China, and 1 in Singapore) via convenience and snowball sampling. 
The inclusion criterion was familiarity with and/or interest in prompting LLMs for daily tasks.
We maintained a balanced distribution of academic backgrounds during recruitment, resulting in 20 participants: 18 students from 9 universities and 2 recent graduates working at 2 companies.
{These participants included 17 males and 3 females, aged between 20 and 30 years.} Among all participants, 12 were from engineering majors, and the others were from {non-engineering backgrounds including music, marketing, biomedical science, business analytics, and statistics}.
Table~\ref{tab:participants} provides more details about participants' background and affiliation, as well as additional information about their general usage of AI/LLM gathered through a pre-study survey.

\subsection{Tasks \& Procedure}
We devised 3 sessions covering different types of LLM-based tasks. 
Every session consisted of a main task, {with some examples of variation in requirements or constraints}. 
\rev{This design aimed to represent cases where users' intent is emergent or exploratory, requiring iterative refinement (usually no more than 5 rounds) of articulation so users' internal goals become more defined through interaction.}

To complete the tasks of each session, we encouraged each participant to {keep iteratively modifying the prompt object until} they felt satisfied with the generated content from LLM, or they could no longer find any further improvements to make.
However, because of the formative nature of this study, we focused on discussing participants' feedback without requiring them to meet a hard requirement for each task.




The specific tasks for each session were as follows.
\begin{itemize}
    \item \textbf{Task 1: Reasoning}. A domain-specific generative task with a known or expected answer, which can be validated by reasoning or ground truth. 
    The content was tailored to participants' academic background: engineering and non-engineering.
    For example, participants with an electrical engineering and computer science (EECS) background completed a {script for training a randomly selected machine-learning model (\eg Logistic Regression, Random Forest, Support Vector Machine, Neural Network)}, whereas participants from social science backgrounds performed a social model derivation or analytical writing task. Participants were asked to compose the entire use case based on their own experience and preferences, and include as many details as possible (e.g., EECS participants were expected to specify requirements such as coding language, library/toolkit, input/output data type and format, etc.). {This task represented scenarios where users already had a clear understanding of the desired output and could anticipate how changes to the prompt would affect the result.}
    
    \item \textbf{Task 2: Creative}. A creative generative task in the form of story writing. Participants began by writing a very general and underspecified {self-defined raw prompt (\eg ``Write a story.'')} to obtain an initial LLM output, then extended or edited the prompt by selecting one or more keywords from a predefined list at each sub-task step. Participants analyzed whether each updated LLM output satisfied the new requirements. This task was designed to cover cases in which users need to convey abstract, perceptual, and imaginative ideas to an LLM.
    
    \item \textbf{Task 3: Investigative}. An open-ended information-searching task, typically in the form of trip planning. This task was mainly user-driven, with minimal guidance and no predetermined ground-truth answer. {Each participant was encouraged to start by choosing one target location of interest} about which they had limited knowledge, then iteratively explore their interests and preferences, mainly relying on information provided by the LLM at each sub-task step, {without using external sources such as Google}. This task aimed to reflect scenarios in which {users might shift their focus and interests when new information is revealed.}
\end{itemize}

During {each task session in this formative study}, we {encouraged participants to assess whether LLM responses were satisfactory and to iterate on prompts to improve response quality.}
We applied think-aloud protocol to let the participants speak out their thoughts {during the prompt refinement process}, including goals, other possibilities they would like to explore, any unsatisfied requirements, and suggestions on features that could be added to the current prototype.

After all tasks were finished, we conducted semi-structured interviews.
We asked participants to compare \oop with the traditional way of prompting in each step and to reflect on advantages and drawbacks. 
We discussed participants' overall attitudes toward \oop, including its usefulness for prompting LLMs, major challenges during the tasks, and suggestions for improvement.

\subsection{Measurement \& Analysis}
We took notes during the think-aloud process in completing the tasks as well as the post-task interviews.
We then conducted a thematic analysis~\cite{thematicanalysis}, segmenting each participant's feedback into atomic points and mapping them to the skeleton of the initial design space.
For each part of the design space, we clustered data points to surface recurring themes, which we summarized into issues that \oop interaction design should address.

\subsection{Findings}
We summarize findings that respond to the two RQs.
Note that, for RQ2, we focus on a high-level overview here, while leaving a detailed presentation of the resultant design space in the following section.

\paragraph{\rev{RQ1: How would users perceive and react to the high-level concept of \oop?}}
Participants demonstrated generally positive attitudes towards the overall concept of \oop. Among all the participants in the formative study, \textbf{14} of them 
strongly felt that the direction which \oop was exploring was meaningful after completing all study tasks using the MVP.
While these participants found little difficulty in understanding the overall concept, some admitted that it may require a high ability of logical thinking to fully utilize \oop. 
Meanwhile, \textbf{5} participants (P4, 7, 11, 16, 19) remained ambivalent, recognizing the potential of \oop but feeling unsure due to the MVP's technical limitation, such as overly simple prompt object structure, responding/processing latency, and LLM hallucination. 
Only \textbf{1} participant (P9) expressed a lack of motivation to use \oop. With a strong background in LLMs and prompt engineering, P9 had already been applying various prompting techniques in daily work and needed no assistance from an external interface/system. As a result, P9 felt that using \oop was unnecessary for their workflow and suggested that \oop might be better targeted at less experienced user groups.

\paragraph{RQ2: How to develop and expand a comprehensive design space for \oop?}
We selected the most commonly-mentioned emergent features from participants' feedback. 
As a result, six recurring themes emerged, which we summarize below and discuss further in the next section as we expand the design space: 


\begin{enumerate}[label=(\arabic*), leftmargin=0.25in] 
    \item \textit{Free-text initialization}. Participants preferred to start with a single rough, free-text prompt (P4, 5, 17, 20) even though it would seem ambiguous at the beginning (P20), suggesting that simultaneously displaying multiple ways of initialization can cause confusion (P8). At the same time, participants found that selecting templates to initialize \oop required specific descriptions and examples to understand (P5, 6), and the current set of templates were too broad, not flexible enough to be helpful (P9, 10, 15). 
    \item \textit{Natural-language property edition}. Participants demonstrated a strong preference to modify the intent property by typing in natural-language (NL) text, instead of being restricted to short property names and values (P1, 2, 3, 6, 14, 15, 16, 20). Furthermore, participants sometimes struggled in summarizing and rephrasing requirements into structured properties (P4, 5, 16, 20), which further emphasized the importance of NL-understanding support. 
    \item \textit{Advanced property structure}. Participants expressed demands for composing more complex, informative properties in task 2 (P3, P10, 14, 15), such as designing the storyline structure (P3, 10), setting up the motif (P3), and developing the hidden plot (P3). Most of them expected a nested structure for describing a property.
    \item \textit{Supplemental actions on properties}. Participants shared varieties of additional ways to manage the properties. Many mentioned an `exclude' action, as they showed stronger awareness of what they did not want than what they preferred (P1, 2, 10, 11, 17). Moreover, participants generally wished to provide more context for the properties to prevent misunderstandings. Meanwhile, most participants felt negligible benefits of suggested different wording candidates to help them describe an intent property and some felt they needed suggestions by examples instead (P5, 9, 10, 12, 14, 20).
    \item \textit{Global refinement beyond individual editing} - Participants showed strong needs for a global refinement across the whole prompt after finishing modifying each individual property. Participants observed problems in LLM response when certain properties were incompatible with the others (\eg a story written in a ``horror'' style but targeting to ``children'')  (P2, 4, 11, 20). Several participants also wished for the system to actively assist in completing the prompt design by actively suggesting more properties that fit the overall context (P10, 11, 13, 16, 17).
    \item \textit{Clarified instruction in final prompt} - Participants observed occasional misunderstandings in LLMs when directly submitting the generated prompt object in a JSON-style format (P1, 2). LLMs might be confused about the target of the prompt, and thus tend to interpret the meaning of the whole prompt object rather than considering it as an instruction. Participants suggested modifying the logic and format for generating final prompts, thus to better clarify how LLM should interpret it (P1, 2, 19, 20).
\end{enumerate}




%% file: 05_design_space.tex
\section{\oop Design Space}
\label{sec:design_space}


\begin{figure*}[!tb]
    \centering
    \includegraphics[width=0.84\textwidth]{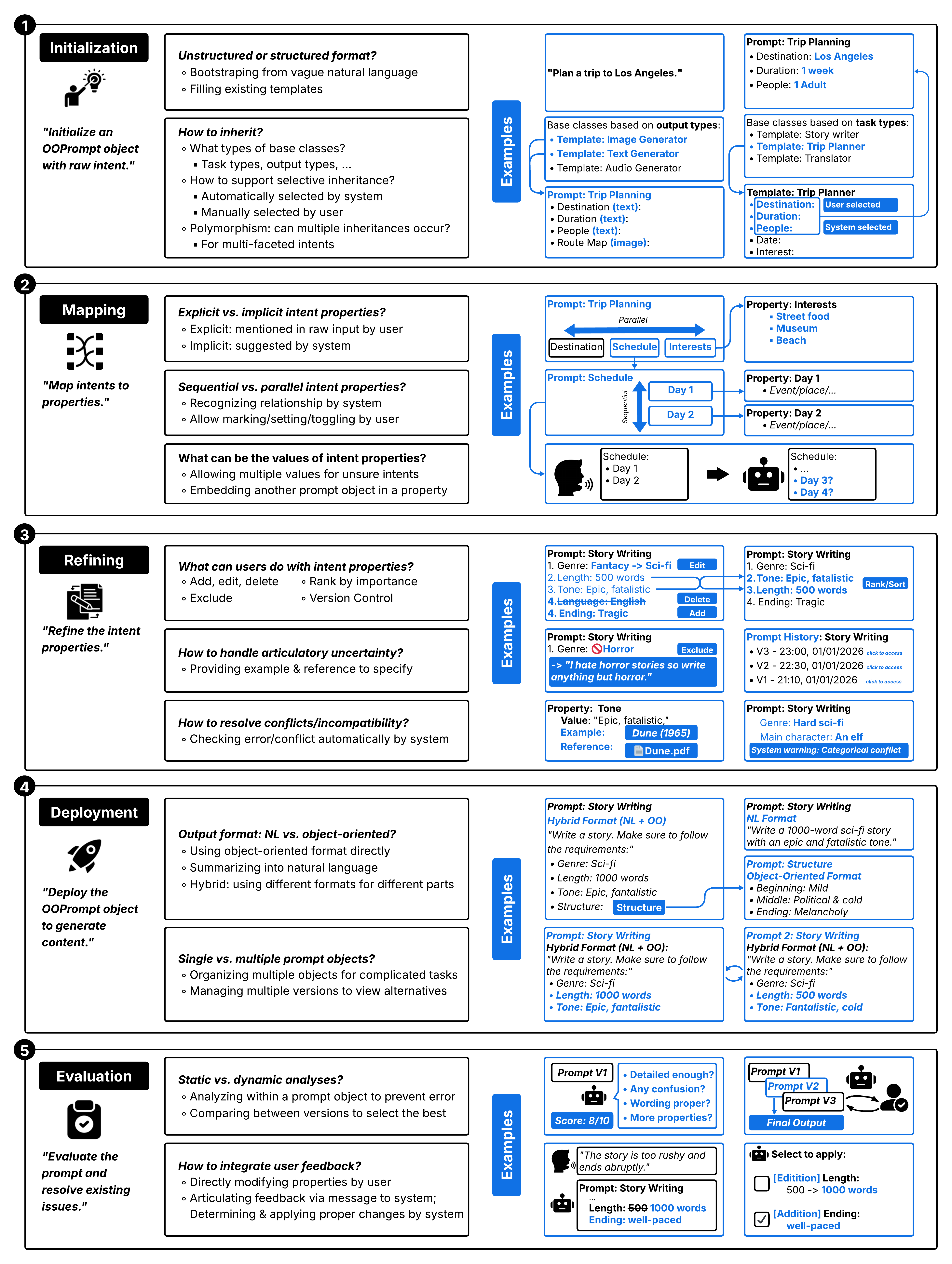}
    \caption{\textbf{The final design space of OOPrompt} updated by insights gathered through the formative study. This illustrates the iterative workflow across five key stages: Initialization, Mapping intents to properties, Refining intent properties, Deployment, and Evaluation. Each stage highlights critical design questions, with example strategies, guiding how users and systems collaboratively structure, manipulate, and optimize prompt objects.
    }
    \label{fig:final-design-space}
\end{figure*}




%
%
%
%
\subsection{Initializing \oop} 
This stage involves users describing their goals or needs as input to formulate an \oop, which may not be fully articulated on the get-go. 

\subsubsection{Unstructured or structured format?}
Should we let the user describe their ``raw'' intents just like how they would input in a regular prompt, or should we let them follow a more structured format, \eg filling in specific fields of ``purpose'', ``task type'', ``audience'', ``expected output criteria''?

The unstructured format should be the default mode because users, at the beginning, might not have thought through the intent properties. Providing structured information fields \inquo{takes some time to think and may be inefficient}6.
P5 mentioned that users with little engineering background may struggle with extracting structured information, but are better at creative and divergent thinking. They usually tend to describe their intent in long complex sentences and expect AI to help them extract the logical structures.

\paragraph{Examples from related work}

{In \textit{Dynamic Prompt Middleware}, users start the process by providing a NL prompt and then entering their goal or query into an open-ended text box~\cite{drosos2024dynamic}. In \textit{Low-code LLM} users first input a task prompt, which is often a very brief description of the task they want to accomplish~\cite{cai2024lowcodellmgraphicaluser} \textit{Luminate} waits for users to provide natural language prompts then uses this input and context to generate dimensions and values~\cite{suh2024luminate}.}

\subsubsection{How to inherit from base classes?}



Besides initializing an \oop by describing one's intent from scratch, it is also possible to inherit a templated \oop from some base classes.
To enable inheritance, an \oop UI should address the following questions.

\paragraph{What types of base classes?}
Admittedly, the small set of base classes used in the formative study is far from enough to meet participants' needs (P10) and the implementation of \oop should include a process of curating base classes relevant to the application context.
Some participants would like to have more fine-grained categories---for example, ``text generator'' is too broad and needs to provide more specific subclasses (P9).
P4 further pointed out that users could benefit from having different templates for individual intent properties as well.
While we currently defined base classes by their output type {(\eg code, plan, story.)}, participants mentioned the need to search for a template by example (P5-6) or by use case (P4-6).


\paragraph{How to support selective inheritance?}
Participants' feedback indicated that \oop UI should support automatically selecting the most appropriate base class while letting the user decide which subset of properties are the most relevant to inherit.
In tasks 2 and 3, some participants (P9-11, 15-16) always deleted some default properties (which were automatically imported from the ``text generator'' template) right after creating the prompt object. They mentioned that those properties might be too general, not needed, or did not exactly match their very use case.

\paragraph{Polymorphism---can multiple inheritances occur?}
One participant mentioned that he \inquo{expect(s) the LLM to return a mix of different types}1.
For example, when generating a poster for advertising an event, a user would want the \oop to inherit from both the text generator and the image generator base classes.
While not a common use case, the \oop UI should nonetheless support users to select multiple base classes to inherit from during initialization.
Further, \oop UI should make it clear that inheriting more than one base classes does not indicate that the intent is ambiguous (P17); rather, it means the intent is multi-faceted and can be decomposed into multiple sub-intents.

\paragraph{Examples from related work}
{\textit{LangGPT} designs the structured artifacts, ``modules'' (similar to ``templates''), based on user role, goal, constraints, and output format. \textit{LangGPT} also applies multiple inheritance by allowing a single prompt to combine multiple modules (Profile, Constraint, Goal, Workflow, Style, etc.)~\cite{wang2024langgptrethinkingstructuredreusable} \textit{Object-Oriented Drawing}'s base classes consist of attributes of digital content (e.g., stroke, fill, drop shadow) and supports multiple inheritance by blending-holding two attribute objects of the same type to generate a child object that takes 50\% from each parent~\cite{oodrawing}}

%
%
%
%
\subsection{Mapping Intents to Properties} 

Mapping intents to properties is necessary whenever there is an unstructured component in the initial user input (\ie the raw intent).
If there is an inheritance of base \oop classes, we should first extract the raw intent to fill in the inherited properties.
Intent properties that do not fit the base classes' will be added separately to the \oop.

\subsubsection{Explicit \vs implicit intent properties?}
Some intents are explicit, corresponding to specific parts of the raw input, \eg the intent of \texttt{``output type''} in the prompt, specified as ``write a short story about ...''
Other intents, however, can be implicit (unspoken), meaning there is no correspondence to the raw input but the user still have them in mind.
For example, a user might intend to generate and tell a bedtime story to their kid but might elude to mention ``bedtime'' in the prompt.
\oop UI should support extracting both types of intent properties.
For implicit intent, P11 mentioned the possibility of ``extrapolating'' from the initial input. For example, if the current (explicit) intent properties include \texttt{``output type: short story''} and \texttt{``topic: animal''}, then the system can use these properties to retrieve other relevant yet unmentioned intent properties, \eg \texttt{``audience''} and \texttt{``occasion''}.

\paragraph{Examples from related work} {One general way is to always start with generic, high-level questions about relevant concepts or facts, then instruct LLMs to rephrase and expand the question to uncover more implicit details~\cite{xiang2025selfsupervised, schulhoff2025promptreportsystematicsurvey}. To this end, \textit{AI-Instruments} captures implicit intents by organizing generated content based on users' preferences and suggests alternatives, supporting an iterative loop where users refine, select, and discard prompts/content~\cite{aiinstruments}. \textit{PromptPilot} leverages the power of AI to proactively suggests task-appropriate prompts and formulates follow-up prompts users might consider, thus supporting intent discovery in a way with reduced cognitive load~\cite{promptpilot}. Moreover, human feedback may also be involved in this process. For example, one may generate multiple responses based on the raw input and letting users to select a preferred one, to systematically refine the prompt to align with the user's implicitly preferred criteria~\cite{lin2024promptopt}.}

\subsubsection{Sequential \vs parallel intent properties?}
Another dimension to distinguish different intent properties is their sequential \vs parallel relationship with each other.
A common example of sequential properties is steps, which is stereotypical when using the chain-of-thought~\cite{wei2023chainofthoughtpromptingelicitsreasoning} prompting techniques.
Multiple participants (P4-5, 10-11) in the formative study used sequential properties in \oop: beginning, events, and ending.
In most other cases, multiple intent properties are considered (more or less) parallel to each other, such as \texttt{narration style} and \texttt{hidden plot} when generating a story (P3).
However, as discussed later, sometimes intent properties are not completely parallel---their values might conflict with each other or they might have compatibility problem.
Further, one participant noted that some prompts might contain sequential properties (\eg the sequence of a story: beginning, events, and ending), yet each of them, in turn, contains more than one parallel properties (\eg style, length, and tone) (P4).
\oop UI should automatically recognize whether certain intent properties are sequential, which should be treated as steps when finalizing the output \oop; further, users should be able to mark or create intent properties as sequential ones, which sometimes might involve grouping multiple parallel properties. {To better organize such complex structures, P20 mentioned the possibility of explicitly toggling between ``parallel'' and ``sequential'' relationship between properties by a button, while P19 suggested to render properties in a table or flowchart to clearly represent all logic relationships. }

\paragraph{Examples from related work} {Sequential intent properties are usually coherent series of intermediate steps~\cite{wei2023chainofthoughtpromptingelicitsreasoning, schulhoff2025promptreportsystematicsurvey}. To identify this type of properties, one most straightforward way is to directly ask users during initialization by questions such as ``Explain your reasoning step by step.''~\cite{schulhoff2025promptreportsystematicsurvey}. It is also possible to instruct LLMs to decompose a problem into sequential sub-problems~\cite{schulhoff2025promptreportsystematicsurvey, khot2023decomposedpromptingmodularapproach, joshi2024coprompter}, or more generally, extract the overall hierarchical structure including both sequential steps and parallel atomic attributes~\cite{kim2023cellsgeneratorsandlenses, joshi2024coprompter, graphologue, shen2025interactionaug}. On this point, beyond one-shot processing, some studies develop iterative frameworks to extract properties in different relationships round by round~\cite{schulhoff2025promptreportsystematicsurvey, ma2023fairnessguidedfewshotpromptinglarge}, while some also apply divide-and-conquer strategy~\cite{zhou2025instructpipegeneratingvisualblocks}.}

\subsubsection{What can be the values of intent properties?}
Other than mapping the types of intent properties, it is also necessary to extract each one's values from the initial raw input.
Conventional properties in objects are often described in text, which is naturally compatible with the textual format of prompts and is the default way of filling in an intent property.
Instead of wordsmithing each intent property in short phrases, some participants suggested the option of typing in free text (P1-2, 6, 14-15).
During the story writing task, instead of editing an intent property, multiple participants opted to send another short prompt to provide additional details to elaborate on an intent. 
LLM's interpretation of such addendum was suboptimal, as its response indicated that more context was needed.
Such existing usage patterns (\ie sending multiple prompts to iterate on one's intent) indicated the need for a more integral approach. 
\oop UI should let users iterate on the same prompt in an object-oriented way, allowing them to freely express each intent property as an embedded component of \oop.
Further, \oop UI can also let users input multiple possible values of an intent property when they feel unsure about the best way to describe it.
In addition, participants mentioned defining an intent property by writing up examples on their own (P9, 10, 14)
or providing external references (P5, 12, 20).
As discussed later, users can resolve such communication gap at deployment time by seeing and selecting different ways of generating the final \oop.
Usually, writing a single, linear prompt in a text box might have limited affordances for users to divide their focus on one specific intent property's examples or references. 
In contrast, for each intent property, \oop UI can support a user to search for, select, and organize one or more examples/references. 

\paragraph{Examples from related work} {Most related projects depend on text value as the easiest but most generalizable type~\cite{wang2024langgptrethinkingstructuredreusable, mao2025promptstemplatessystematicprompt, joshi2024coprompter, khot2023decomposedpromptingmodularapproach, shen2025interactionaug}. Some projects augment on purely text-based values by adding quantization components, usually in numeric values, to further define the extent or strength of certain intent properties~\cite{suh2024luminate, drosos2024dynamic, aiinstruments}. Particularly, \textit{AI-Instruments} supports representing properties by graphical interface objects, such as Fillable Brushes and Transformative Lenses, for transmitting complex, abstract, aesthetic attributes~\cite{aiinstruments}.}


%
%
%
%
\subsection{{Refining Intent Properties}}

Now that a user has initialized an \oop with their intent properties extracted, it is time for them to interact with those properties for further refinement, editing, and iteration.

\subsubsection{What can users do with intent properties?}
Besides the usual operations of addition, edit, and deletion, participants suggested the following actions users can take on intent properties.

\paragraph{Creating hierarchy} 
One intent can be a composite of multiple ``sub-intents'', which enables specifying more abstract or complex intents.
For example, given the default extracted property of \texttt{``output\_type: story''}, a user can create a hierarchy of \texttt{``plot'', ``structure'', ``storyline''} (P10, 14, 15).
As users iterate on their prompts, they might come up with new ``sub-intents'' that can be organized into a proper hierarchy, \eg hidden plot, associated/parallel storyline, echoing to earlier paragraph (P3).

\paragraph{Scoring and/or ranking}
Participants would like the ability to determine the priority relationship among different properties.
Our initial MVP simply arranged intent properties in the order of their appearance in the raw input. 
P1 noted that properties placed at the bottom might receive less attention despite their potential importance. 
Instead, \oop UI should automatically infer or allow users to specify weights for each intent property, based on which the list of intent properties are sorted and ranked.
{Multiple participants, while agreeing on the motivation of scoring and ranking, disliked such methods since it would be too complex for users to determine the weight values or ranking relationships themselves (P1, 2, 17, 18). Some suggested simplifying scoring in numeric values to simple multiple choice on tiers or options (\eg ``not wanted at all, slightly wanted, normal, wanted, highly wanted'') for each property (P17, 18). Specifically, P17 acknowledged the need for automatic sorting by priority or importance, as a user might not have thought about which property is more important than the others. Some participants also supported disabling ranking to avoid explicitly comparison since, it was unnecessary (P1, 2, 17, 18).}

\paragraph{Exclusion}
Interestingly, multiple participants pointed out the need to exclude certain intent properties or certain values of a given intent property.
As one participant put it, \inquo{Users may not know what they want, but know what they don’t}1.
\oop UI should provide a ``do not want'' button next to each extracted intent property (P2, 10, 11, 15, 17, 18).
For a given intent property, users can also specify a ``counter-value'', \eg what not to include in the story's characters.


\paragraph{Version control}
Participants mentioned the need to see change logs and to restore to a specific version of \oop (P6, 12, 15).
P6 suggested a toggle next to each intent property where the user can click to see its historical values.
In addition, participants also would like to see each versions of \oop in the context of the chat history (P2, 12, 17), because how LM responded to previous prompts as well as their ``memory'' might influence how users iterate on the \oop.

\paragraph{Examples from related work}  {Related work show fundamental functionalities of adding, deleting, and editing properties, rendered in different ways including manual editions on text values and visualizations~\cite{suh2024luminate, aiinstruments, mao2025promptstemplatessystematicprompt, drosos2024dynamic, oodrawing, kim2023cellsgeneratorsandlenses, joshi2024coprompter}. Moreover, \textit{Luminate} presents an inspiring design that enables users to change the quantization level of the dimensions by semantic zooming (\eg adding more dots to the columns of certain dimensions on the graphical interface)~\cite{suh2024luminate}. Similarly, \textit{Dynamic Prompt Middleware} allows quantization adjustment for properties by graphical elements such as buttons and knobs~\cite{drosos2024dynamic}. In addition to refining single properties, many studies present examples of storing model configurations to maintain previously promising configurations for versioning~\cite{kim2023cellsgeneratorsandlenses, mishra2025promptaidpromptexplorationperturbation, mao2025promptstemplatessystematicprompt, joshi2024coprompter}. }

\subsubsection{How to handle the uncertainty of intent properties?}
{
\rev{Uncertainty in prompting is often an inherent byproduct of the articulation gap, as human language cannot always perfectly translate internal thoughts and intents into explicit instructions. This form of uncertainty is rooted in the ambiguity of human expression rather than technical factors like model temperature or stochastic randomness. As noted in prior work, Semantic mismatches between user intent and AI interpretation has been identified as a key design challenge~\cite{drosos2024dynamic}}
\rev{Specifically, as a property value usually consists of several key terms, users usually have limited idea how abstractly or specifically the model will generate the content (P4, 5, 9, 12, 16).} For example, by providing the property \texttt{``interest: food''} during a trip planning task, the model could not fully understand if the user expected some specific examples of local food or just wanted a rough overall image of the local eating style (P9). 
Naturally, we came up with approaches such as generating candidates, different versions/forms with the same meaning, to each property value and assigning confidence score by the model to show certainty. However, participants did not find these methods effective. Many participants observed that the appended candidates usually made negligible difference on the generated content (P9, 10, 13, 14). Those against scoring in addressing uncertainty issue stated that scores were not informative in truly guiding users to improve while making the process more complicated (P18, 19). P16 provided another potential alternative about system generating brief explanation for each intent property. Others mentioned further defining an intent property by providing specific examples (P9, 10, 14). Some also mentioned uploading files or documents as external references corresponding to any single intent property (P5, 12, 20).}




\paragraph{Examples from related work} {\textit{AI-Instruments} presented ``Fillable Brushes'', a highly efficient and straightforward way of sampling and referencing---users may ``fill'' an empty brush by using existing content from source images as examples, functioning like a color picker but capturing perceptual, abstract attributes like style~\cite{aiinstruments}. An approach used by multiple related projects involves a large database of specific examples to help users find the right style, which supports refinement of text-to-image prompts by exploring, comparing, and modifying styles through interactive lists~\cite{aiinstruments, shen2025interactionaug, schulhoff2025promptreportsystematicsurvey, kim2023cellsgeneratorsandlenses}. Also, another general method is to let the system provide multiple example options for each attribute/property/dimension, then users may select an option or self-define a new one based on the given options~\cite{aiinstruments, suh2024luminate, drosos2024dynamic, mishra2025promptaidpromptexplorationperturbation, joshi2024coprompter}. }

\subsubsection{How to resolve conflicts/errors/incompatibility?}
One unexpected topic that came up in the formative study was how participants noticed the possibilities of intent properties having conflicts, errors, or incompatibility relative to one another.
This can be a common problem when users iteratively edit a long and complex prompt---some later addition/edition might have conflicts with some earlier elements.
Consider a trip planning example: initially a user might define the length as two days but later change their mind to three days, adding what they want to do on the third day in the prompt, without updating the trip length property. {P2 experienced similar issue in a story writing task; P2 first defined a ``humorous, happy'' tone and a ``good ending'', then changed to ``bad ending'' while forgetting to adjust the tone. This potential conflict resulted weird, low-quality generated content. Therefore, P2 suggested automatic conflict checking among intent properties to prevent users' oversight.}

\paragraph{Examples from related work}  {\textit{Dynamic Prompt Middleware} resolves conflicts between prompt dimensions by dynamically constraining incompatible options (\eg using single-select controls) for conflicting attributes and instructing the model to reconcile or ignore conflicting refinements. Also, \textit{CoPrompter} systematically addresses this issue by decomposing complex prompts into atomic criteria via LLM-based decomposition and metadata analysis. It then highlights overlaps, subjectivity, or inconsistent interpretations among attributes, and presents to users so they may manually update prompt and fix the conflicts.}

\subsection{Deploying \oop}





Once users finish refining an \oop, the next step is to make it ready for deployment, \ie sending it to LLM for generating the desired output.

\subsubsection{Output format: natural language or object-oriented?}
Most prompts by default are sent to LLM in natural language because it is easier for users to compose rather than necessary for LLM to understand.
Therefore, an \oop does not have to be in natural language---in fact, we observed that phrasing multiple intent properties in natural language sometimes resulted in LLM ignoring some properties.
Alternatively, \oop can just present itself in an object-oriented format, such as JSON.
However, context is needed when sending a JSON version of \oop.
P2 observed that sometimes if \textit{only} a JSON object was sent, LLM would not perform the tasks, but try to explain and analyze the meaning of the JSON object. 
Therefore, the recommended approach is helping users generate the JSON-formatted \oop with natural language descriptions, as such a ``hybrid'' format may be more detailed and accurate (P1, 2, 17, 19, 20). P19 agreed that purely a JSON object would not guide LLM well, elaborating that the LLMs need to be well set up with system prompts to solve this issue. Meanwhile, P19 supposed that appending NL instructions to JSON objects might perform similarly to adding a system prompt, so it might be a good strategy.     

\paragraph{Examples from related work}  {
Many projects choose to generate output prompts similar to unstructured, NL prompt, probably with some diction/language refinement and additional sentences carrying extra information~\cite{x3dobject, mishra2025promptaidpromptexplorationperturbation}. Those that import structuring techniques such as templates may support submitting the predefined templates with content inserted to LLMs to generate content~\cite{mao2025promptstemplatessystematicprompt, chainingllmvisual, wang2024langgptrethinkingstructuredreusable, mishra2025promptaidpromptexplorationperturbation}. Specifically, some templates are designed to explicitly differentiate high-level intents and specific details, so they may better guide LLMs to perform exactly the expected tasks~\cite{mishra2025promptaidpromptexplorationperturbation, wang2024langgptrethinkingstructuredreusable}.
}

\subsubsection{Generating single \vs multiple prompt objects?}
In cases where users specify {multiple intent properties that need further elaboration (\eg users may want to assign specific characters and events to each paragraph of a story), \oop UI can also help them generate multiple and/or recursive prompt objects if the initial intent has too complicated structure or is too informative to be encapsulate in a single property value via text articulation (P3, 4, 5, 10, 11, 18, 19)}.   
Another reason to generate multiple prompt objects is to let the user see alternatives---\oop UI can allow them to see different ways of describing their intents other than their edited versions.
As mentioned earlier, users should be able to specify multiple possible values for a given intent property and such uncertainty can be resolved now as the \oop UI synthesize those alternatives into a few different prompt objects. 

\paragraph{Examples from related work} {Based on the idea of CoT~\cite{wei2023chainofthoughtpromptingelicitsreasoning}, related work includes many projects that try to decompose a complicated task, either done by users or systems, into chained sub-tasks for better consistency and performance~\cite{khot2023decomposedpromptingmodularapproach, cai2024lowcodellmgraphicaluser, chainingllmvisual}. Other studies introduce more complex hierarchical structure between the multiple sub-prompt objects. For example, \textit{ChainForge} builds templates that can be recursively nested, thus improving the capability of hierarchy organization~\cite{chainforge}. \textit{AI-Instruments} supports converting sub-objects to instruments that can be reused and even combined with other instruments, allowing even higher flexibility~\cite{aiinstruments}.}


\subsection{Evaluating \oop}

\subsubsection{Static \vs dynamic analyses?}

Similar to programming language, \oop UI can perform some static analysis of the generated prompt object before sending it to LLM.
Note that such an analysis could also happen earlier, as above-mentioned, \eg checking conflicts between intent properties.
In comparison, a final check at this step is more holistic and 
a non-exhaustive list of things to analyze includes 
\one the number of tokens, which might matter if users refer to multiple sources of references when specifying intent properties;
\two similarity with other \oop examples (assuming there is access to a library of them), which can help users see whether they miss any intent properties, \eg forgetting to specify programming language in code generation (P1, 2, 17, 19); and
\three safety, which is a standard measure in general LLM usage, \eg {no dark themes such as ``sorrow'' or ``horror'' for  ``children'' as audience (P4, 20), no instructions on generating explicit/sensitive/toxic content (P11), information legitimacy of any external websites or other types of sources to be included in the output content (P7, 9, 10, 12, 13, 15, 16).}


In addition, \oop UI can let users perform a dynamic analysis by running one or more prompt objects through LLM, which will generate results that are in turn analyzed by another (evaluative) LLM to determine which \oop works better.
First, the user needs to define how the evaluative LLM should judge the results.
Then, the user selects the prompt objects they would like to compare with each other.
Optionally, the user can also select one or more LLMs to ensure a more unbiased perspective.
Next, the \oop UI automatically runs the test and generate a performance report, which can help the user select a version of \oop that are most likely to work.

{Users showed majority preferences on self-evaluation during the study tasks, explaining that though an evaluative LLM may work well for objective tasks such as coding (P17), it cannot guarantee a good evaluation performance on personal, subjective requirements, especially in creative tasks, due to the uncertainty and implicitness in human-system communication (P17, 18, 19). Despite this, for system-side evaluation, participants generally preferred more suggestive evaluation instead of purely scoring or judging (P2, 3, 10, 11, 16, 17, 20).}

\paragraph{Examples from related work} {\textit{CoPrompter} relies on Criteria Questions (CQs) derived from decomposing user input into atomic intents, as well as categorical alignment scores determined by evaluative LLMs, for assessing consistency~\cite{joshi2024coprompter}. \textit{ChainForge} also leverages evaluative LLMs, but to determine evaluation scores based on provided ground truth data~\cite{chainforge}. \textit{Self-Supervised Prompt Optimization} uses a reference-free evaluation paradigm, involving an LLM evaluator performing pairwise comparison of outputs generated by different prompts~\cite{xiang2025selfsupervised}. \textit{iPrOp} involves a human-in-the-loop process and considers four key prompt properties: Performance on annotated data, readability and interpretability, quality of an explanation, and alignment of annotations~\cite{li2025ipropinteractivepromptoptimization}. \textit{Prompt Optimization with Human Feedback} relies entirely on human preference feedback, training a model to predict a latent score for each prompt~\cite{lin2024promptopt}. \rev{\textit{EvalLM} uses an LLM-as-a-judge to provide feedback based on user-defined, qualitative criteria~\cite{kim2024evallm}.}}

\subsubsection{How to enable and integrate user feedback?}
Other than the above-mentioned systemic analyses, users should also be able to directly edit or rank the prompt objects, such as rephrasing an intent property inline or mark preferences of certain versions of the \oop for later references (P1, 2, 3, 4, 17, 18). 
Further, users can also provide feedback, via free text input, after reviewing the \oop holistically (P2, 6, 15, 20).
Instead of going back to editing low-level intent properties, \oop UI should recognize and incorporate such holistic feedback and automatically update necessary intent properties for the user. P19 also mentioned the possibility of iteratively adjusting criteria of the evaluative LLM, instead of prompts, based on human feedback. 

\paragraph{Examples from related work} {\textit{PromptPilot} waits for users to share some keywords or phrases and directly generates relevant prompts tailored to the task context~\cite{promptpilot}. \textit{PromptAid} allows users to select system recommendations for keyword suggestions or paraphrases from a recommendation panel then  test the custom-curated examples with LLMs to see the effects~\cite{mishra2025promptaidpromptexplorationperturbation}. \textit{Dynamic Prompt Middleware} enables users to generate their own new options via NL prompting, thus having more graphical elements on the interface to be interactively adjusted~\cite{drosos2024dynamic}.}

%% file: 06_validation.tex

\section{Validation Study}
\label{sec:validation_study}

Guided by the design space, we iterated on the MVP to develop a full prototype that addressed the main design questions throughout the life cycle of an \oop.
We then conducted a validation study to 
\rev{further understand the effects of \oop on users' interaction with LLM. Specifically, we are interested in answering the research question:}
{
\begin{itemize}
\item[RQ3] What are the added values and trade-offs of using \oop compared with traditional prompting approaches?
\end{itemize}
}

\subsection{Full \oop Prototype}

\begin{figure*}[!tb]
    \centering
    \includegraphics[width=1.0\textwidth]{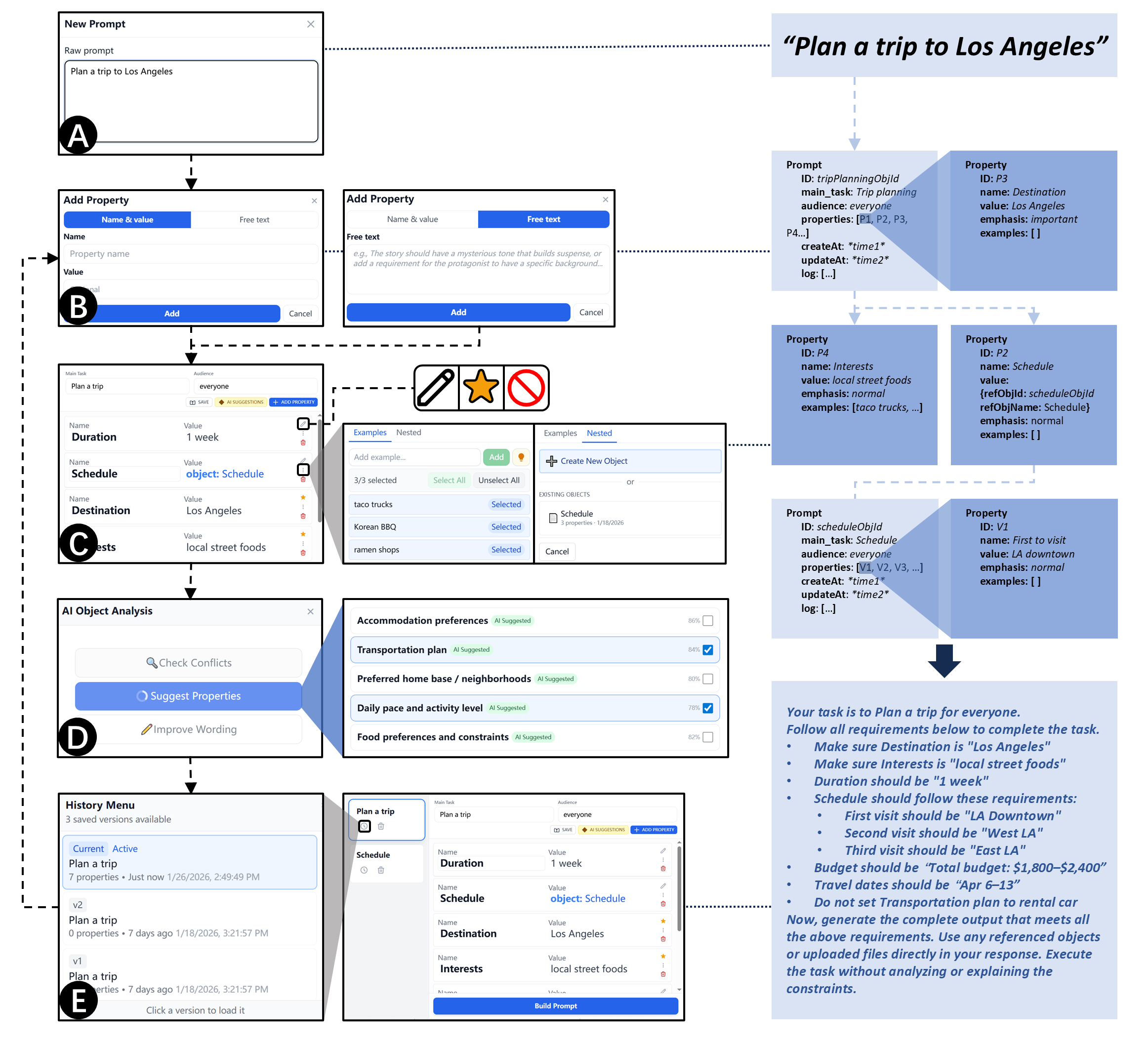}
    \caption{\rev{The \oop system pipeline and underlying data structure. \textbf{(A)} Initialization: A raw natural language intent is reified into a base Prompt Object. \textbf{(B)} Property Definition: Abstract attributes are encapsulated into Property Objects with specific states like emphasis and examples. \textbf{(C)} Hierarchical Structure: Users manage complexity through nested organization, where a property can instantiate a child Prompt Object, creating a parent-child relationship. \textbf{(D)} Evaluation \& Suggestion: The system provides AI-assisted evaluations with suggested solutions (\eg more properties, conflict check, language refinement). \textbf{(E)} Versioning \& Deployment: The final prompt is generated via a hybrid logic approach, while previous versions are stored as distinct, restorable assets in a history menu.}}
    \label{fig:prototype_ui}
\end{figure*}

\paragraph{\rev{Prototype Interface \& Walkthrough Example}}
\rev{We implemented a full \oop prototype to operationalize the complete design space and conduct validation study.
Fig.~\ref{fig:prototype_ui} 
provides a technical walkthrough of the \oop prototype, illustrating the correspondence between the graphical user interface and the underlying hierarchical prompt objects. To illustrate the system in practice, consider a user planning a multi-city travel itinerary. The workflow begins when the user enters a broad intent, such as ``Plan a trip to Los Angeles,'' which the system reifies into an initial Prompt Object (\textbf{Fig.~\ref{fig:prototype_ui}A}). To refine this, the user adds specific Property Objects like ``Destination'' and ``Interests'' (\textbf{Fig.~\ref{fig:prototype_ui}B}). The user may manually type in or call the AI to generate examples for any property, in case that a text value is not enough to clearly articulate the intent (\textbf{Fig.~\ref{fig:prototype_ui}B}).
For example, when the user is interested in ``Local street food'' during the trip, they can specify it to be similar to ``taco trucks'' or ``BBQ''. 
For complex requirements like a daily ``Schedule,'' the user utilizes Nested Organization (Fig.~\ref{fig:prototype_ui}C) to instantiate a child Prompt object, effectively breaking a large task into manageable, hierarchical sub-tasks. After finishing one round of Prompt Object construction, the user may trigger AI Analysis (Fig.~\ref{fig:prototype_ui}D) to analyze and generate improvement suggestions (\eg suggesting more properties, such as ``Daily pace'' to ensuring the prompt is comprehensive). Finally, the user may export an output prompt from the \oop Object and send to any preferred LLM for content generation, while being able to toggle between different saved versions in the History Menu (Fig.~\ref{fig:prototype_ui}E) to compare outputs.}

\subsection{Study Procedures}

We recruited 8 returning participants from the formative study. 
\rev{Running through the whole test without any supervision or assistance from our team, participants were first provided with a six-slide tutorial with graphical illustrations that demonstrate how the functionalities work via a complete walkthrough example. Our goal was to ensure that participants were adequately familiar with the prototype, thus preventing low performance caused by under-utilization of features.}

\begin{figure*}[!tb]
    \centering
    \includegraphics[width=0.95\textwidth]{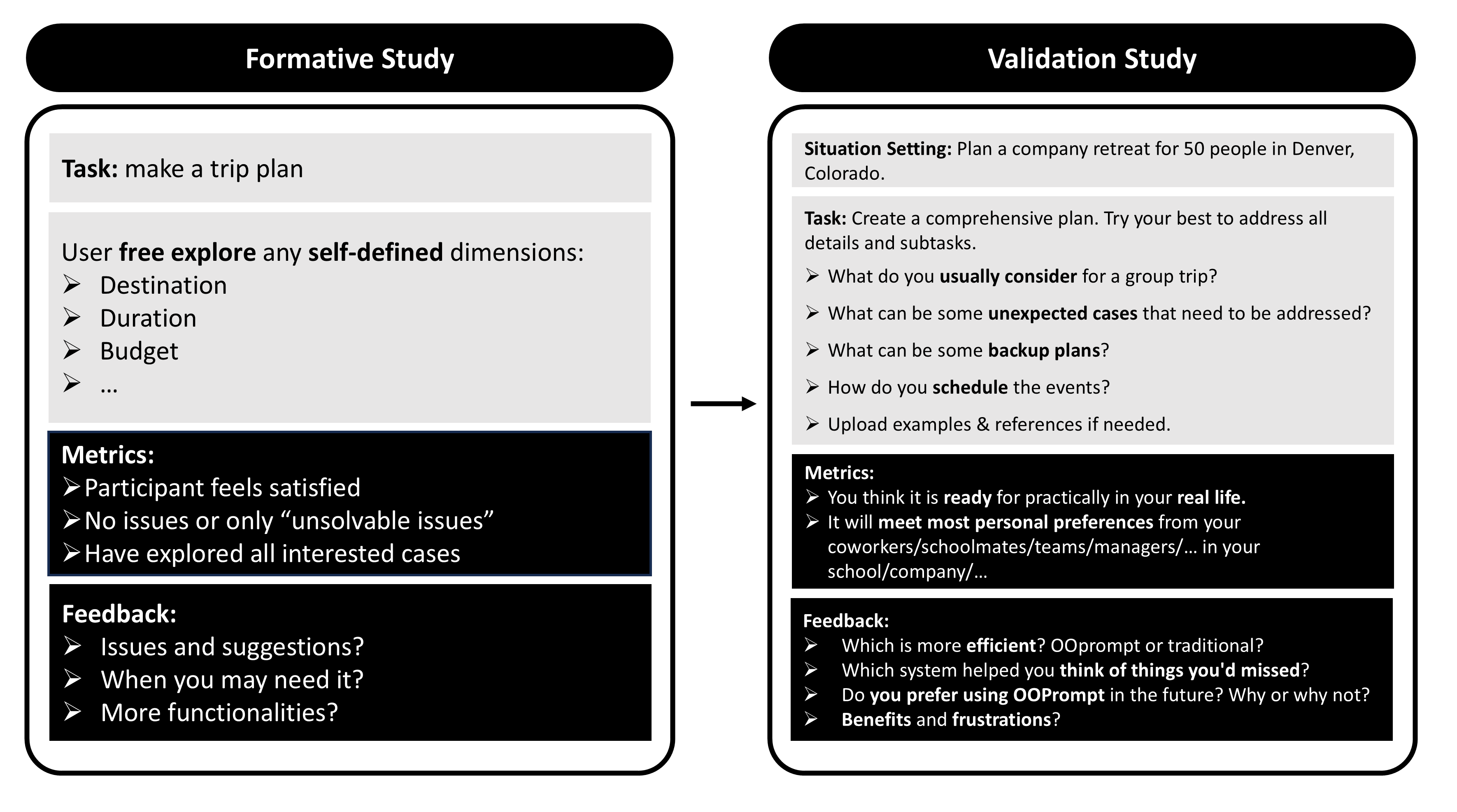}
    \caption{{\textbf{An example showing the differences between the task settings in formative study and validation study}. The formative study focused on free exploration of users' self-defined goals to uncover deficiencies and gather suggestions. However, the validation study described specific scenarios and provided well-defined task requirements for users to accomplish.}}
    \label{fig:formative_task_vs_validation_task}
\end{figure*}

While the formative study consisted of mostly broad, unspecific tasks to stimulate open-ended exploration and to seek constructive suggestions, the final validation study involved well-defined, real-world scenarios, with clearly specified target audiences and explicit evaluation criteria. 
Each scenario is presented following a uniform format: {Scenario Setting}, {General Requirements}, and multiple {Tasks} with guide. Fig.~\ref{fig:formative_task_vs_validation_task} illustrated one example.

\rev{Before the validation study began, we first ensured that (1) each participant fully understood the purpose and meaning of all task descriptions, and (2) each participant was capable of empirically assessing whether the generated content was in sufficient quality, based on their relevant real-life experience.} 
Under this condition, we defined the primary criterion for task completion as ``producing content that was ready for practical use''. In other words, the generated content should be able to address any real-world considerations and personal demands that the participant could identify. 
\rev{This setup aimed to elicit the most accurate and contextually grounded feedback on how well \oop supported the specific needs from participants with different background. }


During the validation study, we provided three scenarios that were intentionally designed to represent different levels of constraint density and complexity. In each scenario, participants were first asked to try traditional text-based prompting (which was inherently supported by the prototype), and then to perform the task with \oop.
\rev{We did not counterbalance the order of \oop~ \vs traditional prompting because using \oop first would likely have a larger carryover effect (\ie participants would already know what a prompt should contain after using \oop).} Table~\ref{tab:study_scenarios} presents details of the scenario settings, tasks, and key considerations, as well as the types of cases they target.

\begin{table*}[t]
\centering
\color{\revcolor}
\caption{\rev{\textbf{Validation study test scenarios with the cases they targeted to evaluate, settings, and tasks.}}}
\label{tab:study_scenarios}
\small 
\begin{tabularx}{\textwidth}{@{} >{\raggedright\arraybackslash\bfseries}p{0.14\textwidth} >{\raggedright\arraybackslash}p{0.18\textwidth} X X @{} }
\toprule
\textbf{Scenario} & \textbf{Targeted Case} & \textbf{Setting} & \textbf{Tasks} \\ \midrule

1. Technical Reporting & High constraint density & 
\begin{tightitemize}
    \item \textbf{Setting:} A project is two weeks behind schedule. Report to different audiences.
    \item \textbf{Considerations:} Tone, communicative purpose, focus, etc.
\end{tightitemize} & 
\begin{tightitemize}
    \item \textbf{Task 1.1:} Report to peer group.
    \item \textbf{Task 1.2:} Report to manager.
    \item \textbf{Task 1.3:} Report to angry stakeholder.
\end{tightitemize} \\ \midrule

2. Group Trip Planning & High hierarchical depth & 
\begin{tightitemize}
    \item \textbf{Setting:} Retreat for 50 people in Denver, Colorado. Plan the trip.
    \item \textbf{Considerations:} Unexpected cases, backup plans, event scheduling, etc.
\end{tightitemize} & 
\begin{tightitemize}
    \item \textbf{Task 2:} Make a plan. Address all details for any potential subtask. 
\end{tightitemize} \\ \midrule

3. Purposeful Debate & High structural uncertainty, low compressibility & 
\begin{tightitemize}
    \item \textbf{Setting:} Alex prefers stocks; Bill wants a world trip. Navigate this conflict.
    \item \textbf{Considerations:} Reasoning integrity, persuasiveness, etc. 
\end{tightitemize} & 
\begin{tightitemize}
    \item \textbf{Task 3.1:} Support Alex against Bill.
    \item \textbf{Task 3.2:} Support Bill against Alex.
    \item \textbf{Task 3.3:} Support yourself against both.
\end{tightitemize} \\ \bottomrule
\end{tabularx}
\end{table*}

\subsection{Data Collection \& Analysis}
After finishing the tasks in all scenarios, we conducted a semi-structured interview with each participant, using the following questions to elicit their feedback on \oop:
\begin{enumerate}
    \item Jointly considering the effectiveness and efforts of prompting, how would you compare \oop with traditional prompting? Which approach would you prefer for each of the three scenarios? 
    \item Which prompting approach helped capture considerations that you would have missed?
    \item If you need to perform tasks similar to the ones in this study regularly at a daily basis, which prompting approach would you prefer?
    \item What types of real-world task would benefit most from \oop?
    \item What frustrated you about \oop? What are the current limitations of the prototype?
\end{enumerate}

We took a thematic approach~\cite{thematicanalysis} 
to analyze the qualitative data from participants' free responses. First, we broke down the transcribed text into atomic segments. We then clustered and coded the extracted segments, and finally reviewed the organized feedback to summarize participants' perceived efficiency, effectiveness, and prompting preferences when using \oop across different task scenarios. We further examined participants’ choices and corresponding explanations to identify trade-offs and their rationales. The findings were synthesized to characterize the main dimensions of these trade-offs and to reveal how users interpreted the advantages and limitations of \oop compared with traditional prompting practices.


\subsection{Findings}

Our findings reveal both the added values and trade-offs of \oop.

\textbf{By turning prompts into modular, editable properties, \oop makes planning and iterative writing faster and clearer than traditional prompting, and better supports ambiguous, exploratory tasks.}
We observed several prominent benefits of \oop. In Scenario 1 (evolving report writing), five of seven participants considered \oop more efficient, while one preferred traditional prompting and two perceived no difference. In Scenario 2 (event planning discovery), seven participants preferred \oop, with one reporting no difference.
In the first two scenarios, most participants identified certain task types that best showcased the strengths of \oop, such as planning with decision making (U2, U4, U6) and report drafting involving iteratively rewriting (U3, U5, U6). \rev{Specifically, U2 described that \oop was preferred for \dirquo{tasks that require planning, tone change, or anything that is actually creation but not question-answering.} Meanwhile, U7 tried to summarize from high level that \oop would be generally helpful with \dirquo{tasks with clear patterns.}
Compared with traditional prompting, participants tended to describe \oop as offering a clearer conceptual structure and helping them organize thoughts and task components before generating the final prompt.
This highlighted {\oop}’s capability in modular representation, which provided a reusable medium for iterative modifications and thereby improved the efficiency and consistency of prompt evolution (U3, U5, U7).}
Moreover, some participants pointed out {\oop}’s usefulness in handling more broadly, ambiguously defined instructions. \rev{Participants generally agreed that \oop helped users capture and articulate concepts they might have otherwise overlooked (U1–U8).} 
\rev{Participants also found \oop better handle tasks that were difficult to grasp (U1, U8) during prompt construction. For instance, U8 mentioned that \oop would be useful when \dirquo{users themselves do not completely understand what the task is referring to;}
U1 stated another advantageous case for \oop was that \dirquo{users do not have clear ideas in mind on how they want the result to be.}} All the feedback collectively suggested that \oop enables more deliberate manipulation and exploration over the logic and completeness of the task framing.


\textbf{Meanwhile, \oop struggles with opinion-forming reasoning and it introduces extra steps that add latency, disrupt chat continuity, and need more automation to streamline its integration into human-LLM interaction.}
In Scenario 3 (debating and convincing), 
one major issue was that participants found it difficult to convert reasoning and argumentative logic into structured intent properties, which led them to continue composing most inputs in long natural language sentences even when using \oop (U2, U3, U7). As a result, \oop provided limited improvement in prompting efficiency in this scenario, highlighting its weakness in efficiently capturing and expressing nuances in users’ reasoning processes.
Beyond this specific scenario, some participants also experienced noticeable processing latency during different steps, including calling AI suggestions and generating the final prompt for deployment (U1, U3, U5). This might have been caused by the serially connected LLM assistants in the workflow structure.
Although this technical issue could be improved with faster, lighter-weight models in the future, it still suggests how \oop would inevitably introduce extra system processing overhead and slow down the back-and-forth interaction between users and LLM.
Several participants also noted that while conventional prompting involved continuous dialogue contexts for human–LLM conversation, \oop treated prompts as relatively independent query objects, making it harder for them to quickly adapt to this new interaction paradigm (U1, U4, U7, U8). Furthermore, U2 specifically expressed the demand for higher automation by describing that the property suggestion module or other AI objective analysis module could present available results as options immediately after raw-text initialization, instead of waiting for users to click on buttons to call the wanted functionalities.

\rev{Based on findings in the validation study, we summarize values and trade-offs brought by \oop, compared to traditional prompting, in Table \ref{tab:value-added}. 
The table highlights how shifting to a modular, object-oriented structure enhances efficiency and organization.}
In the next section, we further discuss the underlying reasons of the pros \& cons from both conceptual design and implementation perspectives. We also propose potential strategies to address the current drawbacks, leaving their experimental validation and the exploration of new optimization approaches for future work.

\begin{table*}[t]
\centering
\small
\color{\revcolor}
\caption{\rev{\textbf{Structured summary of where \oop adds value and where trade-offs appear}, based on participant feedback across scenarios.}}
\begin{tabularx}{\textwidth}{@{} L{0.13\textwidth} L{0.22\textwidth} L{0.22\textwidth} Z @{} }
\toprule
\textbf{Capability} & \textbf{Traditional Prompting} & \textbf{\oop} & \textbf{\oop's Value \& Trade-offs} \\ 
\midrule

\textbf{Representation} &
Linear, monolithic text strings. &
Modular \emph{intent properties} edited in isolation. &
\textbf{+} Clearer structure; safer iterative edits and branching. \newline
\textbf{--} Up-front setup can feel like overhead for tiny tasks. \\ \addlinespace

\textbf{Organization} &
``Wall of text'' with implicit hierarchy. &
Hierarchical composition of sub-tasks/objects. &
\textbf{+} Granular decomposition without losing global context. \newline
\textbf{--} May over-structure when goals are fluid or ill-formed. \\ \addlinespace

\textbf{Intent Articulation} &
Intent remains implicit; easy to miss constraints. &
System-assisted property suggestions/extraction. &
\textbf{+} Surfaces overlooked concepts and gaps. \newline
\textbf{--} Users wanted more proactive, auto-filled suggestions; quality varies by model. \\ \addlinespace

\textbf{Iteration Flow} &
Conversational back-and-forth in a single chat thread. &
Object-centred editing with history and versioning. &
\textbf{+} Controlled differences; reproducibility across versions. \newline
\textbf{--} Can fragment the ``chat rhythm'' some users prefer. \\ \addlinespace

\textbf{Execution Pipeline} &
Typically one-shot calls. &
Multi-step pipeline. &
\textbf{+} Better auditability and reuse. \newline
\textbf{--} Added latency and token cost from serial steps. \\ \addlinespace



\textbf{Evaluation} &
Manual check in chat by human users. &
LLM-assisted analysis (\eg length, style, cohesiveness, etc.) &
\textbf{+} Faster, more consistent comparisons. \newline
\textbf{--} Metrics may miss nuance; needs human review for judgment tasks. \\ 

\bottomrule
\end{tabularx}
\label{tab:value-added}
\end{table*}

%% file: 07_discussion.tex
\section{Discussion}
\label{sec:discussion}


In this section, we discuss possible reasons behind the observed strengths and trade-offs of \oop, aiming to uncover the fundamental trade-offs between \oop and traditional prompting from both conceptual and implementation perspectives.
Then, we discuss limitations of this research and future work.

\subsection{\oop Values \& Trade-offs}

\subsubsection{\oop excels at tasks with
high constraint density}
Tasks with high constraint density, such as report or document writing, typically involve multiple elements (e.g., tone, audience, language style, and focus) that must be integrated coherently through prompting. In such cases, users often compose several descriptive sentences in parallel, each addressing one particular aspect. {\oop}'s modular structure simplifies this process by allowing users to formalize and organize these elements systematically. The advantage becomes especially marked when the task includes numerous requirements that span different dimensions. {However, for tasks with low compositional demand, prompts in a structured format may even negatively affect overall performance. Transforming a simple prompt with only a few conditions into an \oop structure does not yield the expected advantages, such as reduced effort in sentence formulation, easier iterative modification, or clearer logical organization. Instead, this transformation can introduce unnecessary complexity and potentially disrupt the natural continuity of the original prompt.}

\subsubsection{\oop outperforms in tasks with high hierarchical depth}
We observed that a task's hierarchical depth can affect {\oop}'s effectiveness. This is reflected in the degree to which a task can be decomposed into multiple sub-levels that follow a clear abstraction hierarchy. Tasks with explicit hierarchical organization align well with {\oop}'s nested property mechanism. For example, a story may involve several layers, ranging from the main narrative event (top layer) to detailed scene-level attributes such as characters’ clothing color or hairstyle (bottom layer). This structure enables users to extract and manage the task’s logical organization across abstraction levels, facilitating construction and refinement of \oop objects at specific levels as needed. By contrast, when task requirements have no further hierarchical details to extract, prompting in \oop may feel unnecessary or redundant because it may not produce meaningful differences from a raw prompt.

\subsubsection{\oop specializes in tasks with strong repetitiveness}
Sometimes a task or its sub-components are repeated, refined, or reused across similar contexts. Tasks with high repetitiveness, such as routine report updates or iterative content revisions, tend to benefit from {\oop}'s modular and reusable structure. By encapsulating each intent or property as an independent object, \oop allows users to adjust or regenerate specific parts without rephrasing the entire prompt or recapping previously mentioned contextual information, thereby improving efficiency and consistency across iterations. In contrast, highly unique or one-off tasks offer fewer opportunities for reuse, diminishing the advantages of {\oop}'s object-based representation.

\subsubsection{\rev{\oop struggles with tasks where constraints are highly incompressible}}
\rev{There are varying degrees of compressibility when specifying intermediate reasoning or detailed instructions during human-LLM communication.} For example, when prompting an LLM to generate descriptive text with a desired effect, users may choose to provide only a few target keywords (\eg beautiful, delicious, hot) and allow the model to determine how to produce these effects, or they may explicitly prescribe specific rhetorical techniques to be applied in writing. When users care deeply about the concrete methods or reasoning steps through which the task is executed, \oop can substantially reduce the time and effort required to encode such details in the prompt. However, \oop becomes less useful when no compression or abstraction is possible, in other words, when users must explicitly specify every reasoning step or detail to achieve correctness. For instance, in tasks like logical argumentation with fixed reasoning logic, users must input the full reasoning chain as plain text regardless of prompting method. In this case, \oop offers little improvement in efficiency while adding time and cognitive cost due to the interactive modules.

\subsection{Limitation \& Future Work}
\subsubsection{Response speed and automaticity}
\begin{figure}
    \centering
    \includegraphics[width=1.0\linewidth]{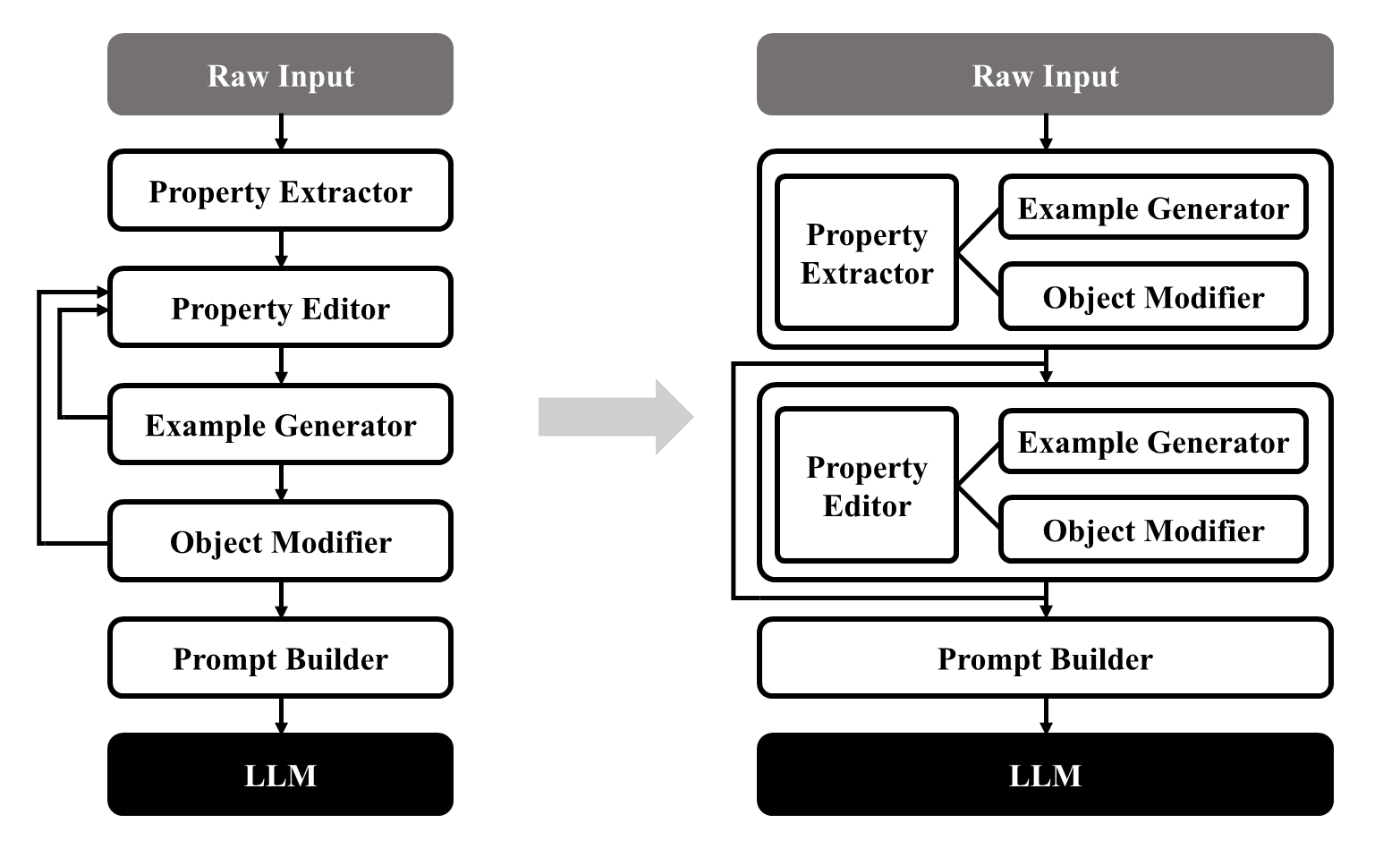}
    \caption{{\textbf{Example implementation of an improved \oop workflow} that autonomously \textbf{calls multiple LLM assistants in parallel}. After the user submits the raw input, the system simultaneously extracts properties and generates corresponding examples and modification suggestions within the same step. All results are produced automatically without additional user activation and are presented as options for selective application to the prompt object. The same procedure repeats for each newly confirmed version of the property space, reducing response latency and improving overall workflow automation.}}
    \label{fig:serial_to_parallel}
\end{figure}

{The relatively high response latency mainly stems from the current prototype’s workflow structure, which involves multiple LLM assistants, each responsible for a specific module. During a full \oop lifecycle, users may need to query these assistants sequentially through multiple API calls, causing accumulated waiting time and reduced overall processing speed. To address this issue, we consider it together with users' suggestions for enhancing automation. Fig.~\ref{fig:serial_to_parallel} illustrates one updated design. Instead of querying assistants in strict serial order, future \oop workflows may improve by allowing parallel calls to multiple assistants within a single stage. This design would significantly reduce total response time and create a more automatic user experience. However, careful tuning of multithreading settings and assistant parameters will be required to ensure a smooth, balanced experience and prevent overly redundant or constrained options at each stage.}

\subsubsection{\rev{Structured tasks with uncertain complexity and low compressibility}} \rev{We believe that the difficulties encountered by {\oop} in tasks with high structural uncertainty are primarily due to the nature of the ``OO'' paradigm rather than inefficient implementation. In tasks such as philosophical debate or sequential reasoning chains, an instruction's value usually lies in holistic flow rather than modular units. While \oop is optimized for encapsulation and reusability, it may lead to friction when users must decompose incompressible, monolithic logic into discrete, hierarchical properties. To mitigate this, participant U3 suggested including a \dirquo{background setting as a default property} to better support unorganized intent. Integrating a ``Global Context'' as a default primary property in the base class could help the system better handle unstructured logic while maintaining the benefits of the OO framework. Future work should explore a pipeline that uses this context layer to facilitate an ``unstructured-to-structured'' transition, allowing the system to generalize more effectively across task types.}

\subsubsection{Dialogue history context}
{Some participants raised concerns about dialogue-history discontinuity, which likely reflects a fundamental design issue of \oop. In traditional prompting, users typically add supplementary information through follow-up queries within the same dialogue context~\cite{sarkar2025conversational, liu2025user}, rather than rewriting the entire prompt to incorporate previous and new content. In contrast, \oop automatically integrates full preceding context into each updated prompt, essentially performing what users often avoid in traditional prompting, albeit implicitly. To prevent inconsistencies between earlier and later prompt versions (where information may conflict or shift in meaning across coexisting versions), our current prototype intentionally separates each query from full dialogue history. During the validation study, participant U8 suggested a potentially feasible solution: converting early dialogue context into static properties that are consistently included in prompt objects. While this approach could help maintain continuity without sacrificing consistency, efficient design of the required data structures and transformation mechanisms remains an open question for future work.}

\begin{figure}
    \centering
    \includegraphics[width=1.0\linewidth]{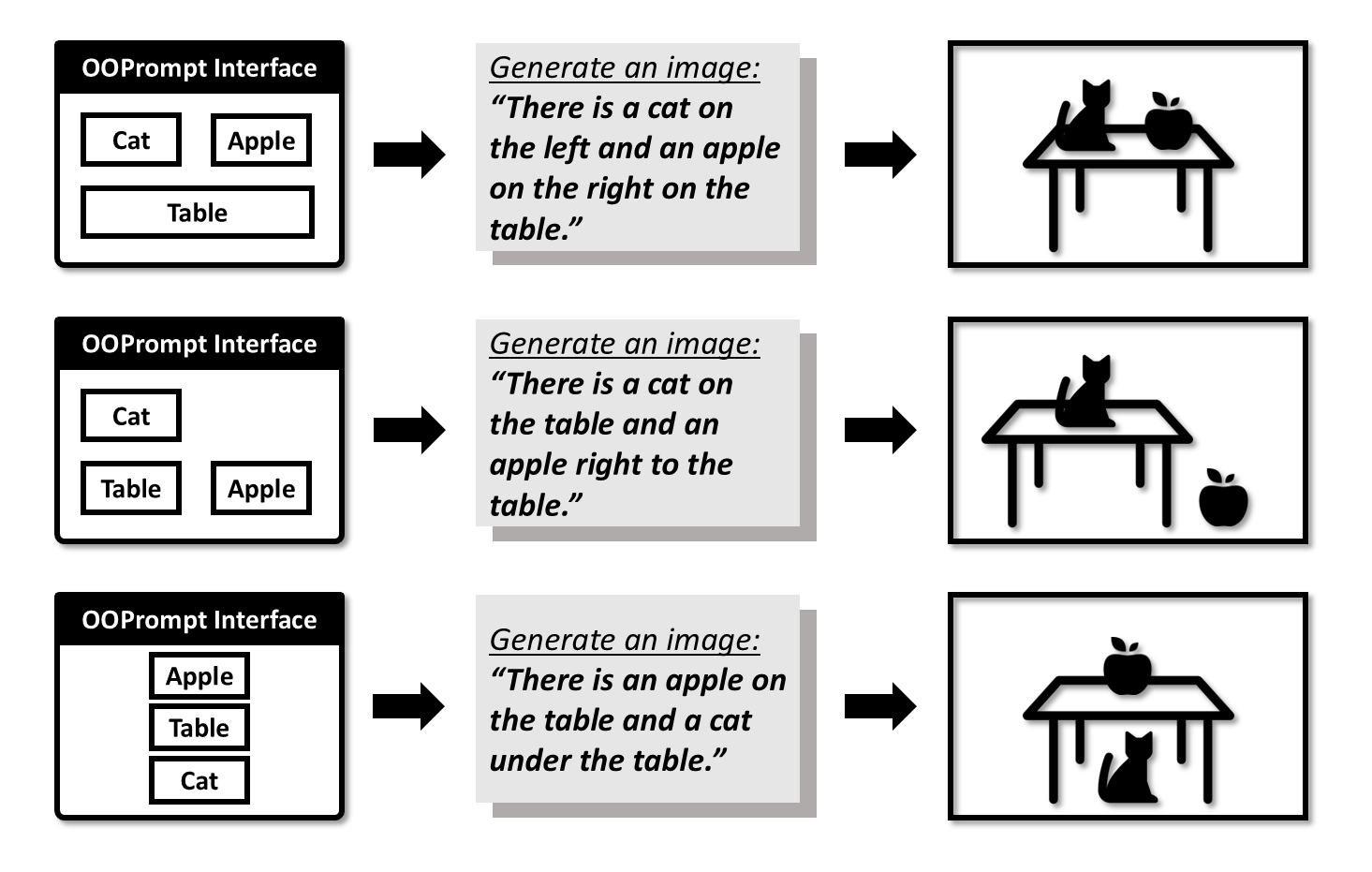}
    \caption{\textbf{\mrev{A schematic illustration of \oop's envisioned spatial manipulation of intent properties in image generation tasks.}} By arranging and placing property blocks of cat, apple, and table within the \oop interface, users can simply control spatial relationships of these three objects in the generated image.}
    \label{fig:position_control_image_generation}
\end{figure}

\subsubsection{Multi-modal task support}
{Although we initiated preliminary discussions on extending \oop to multi-modal tasks (\eg generating content that combines text and images), the current prototype evaluated in the validation study supports only text-based tasks such as writing and coding. This serves as an initial step toward exploring the broader potential of \oop as a prompting paradigm. We believe that incorporating additional modalities (\eg images, audio, or video) could inspire new directions for optimization and generalization.
\rev{For example, modern state-of-the-art text-to-image models like DALL-E~\cite{ramesh2022, betker2023} still lack a formal, user-facing structural paradigm for manipulating discrete visual entities. 
Meanwhile, techniques like ControlNet~\cite{zhang2023adding} have introduced spatial constraints to guide the text-to-image generation process. 
We see {\oop}'s potential to be a bridge to provide users with explicit, object-level control over the prompt, as well as a more simplified, semantic way of spatial and structural guidance.}
Future work may investigate how abstract representations from different modalities can be efficiently utilized to enrich the \oop framework. 
For instance, Fig.~\ref{fig:position_control_image_generation} shows a simple visual interface design of representing intent properties through spatial cues (e.g., vertical or horizontal placement) could potentially be mapped to compositional attributes in image generation, such as the positioning of main objects or characters.}

\section{Conclusion}
In this work, we build on an emerging body of research to formalize \oop as an interaction paradigm in which users compose prompts as structured and manipulable interactive objects. Through a formative study, we collected participant feedback to construct the \oop design space, which was subsequently implemented as a functional prototype and validated in a second study. \rev{The findings suggest that while \oop enhances clarity and efficiency in iterative, exploratory tasks through modular structure, it struggles with subjective opinion-forming and introduces extra steps that require greater automation to truly streamline the human-LLM workflow.}  \rev{Our work contributes theoretical and empirical guidance to the design and engineering of prompt-based, LLM-enabled interactive systems.}

%% file: references.bib
@inproceedings{suh2024luminate,
author = {Suh, Sangho and Chen, Meng and Min, Bryan and Li, Toby Jia-Jun and Xia, Haijun},
title = {Luminate: Structured Generation and Exploration of Design Space with Large Language Models for Human-AI Co-Creation},
year = {2024},
isbn = {9798400703300},
publisher = {Association for Computing Machinery},
address = {New York, NY, USA},
url = {https://doi.org/10.1145/3613904.3642400},
doi = {10.1145/3613904.3642400},
booktitle = {Proceedings of the 2024 CHI Conference on Human Factors in Computing Systems},
articleno = {644},
numpages = {26},
location = {Honolulu, HI, USA},
series = {CHI '24}
}

@inproceedings{aiinstruments, series={CHI ’25},
   title={AI-Instruments: Embodying Prompts as Instruments to Abstract \& Reflect Graphical Interface Commands as General-Purpose Tools},
   url={http://dx.doi.org/10.1145/3706598.3714259},
   DOI={10.1145/3706598.3714259},
   booktitle={Proceedings of the 2025 CHI Conference on Human Factors in Computing Systems},
   publisher={ACM},
   author={Riche, Nathalie and Offenwanger, Anna and Gmeiner, Frederic and Brown, David and Romat, Hugo and Pahud, Michel and Marquardt, Nicolai and Inkpen, Kori and Hinckley, Ken},
   year={2025},
   month=apr, pages={1–18},
   collection={CHI ’25} }

@misc{drosos2024dynamic,
      title={Dynamic Prompt Middleware: Contextual Prompt Refinement Controls for Comprehension Tasks}, 
      author={Ian Drosos and Jack Williams and Advait Sarkar and Nicholas Wilson},
      year={2024},
      eprint={2412.02357},
      archivePrefix={arXiv},
      primaryClass={cs.HC},
      url={https://arxiv.org/abs/2412.02357}, 
}

@inproceedings{Liu2024structuredoutput, series={CHI ’24},
   title={“We Need Structured Output”: Towards User-centered Constraints on Large Language Model Output},
   url={http://dx.doi.org/10.1145/3613905.3650756},
   DOI={10.1145/3613905.3650756},
   booktitle={Extended Abstracts of the CHI Conference on Human Factors in Computing Systems},
   publisher={ACM},
   author={Liu, Michael Xieyang and Liu, Frederick and Fiannaca, Alexander J. and Koo, Terry and Dixon, Lucas and Terry, Michael and Cai, Carrie J.},
   year={2024},
   month=may, pages={1–9},
   collection={CHI ’24} }

@article{promptpilot,
author = {Soomin Kim and Jinsu Eun and Yoobin Elyson Park and Kwangwon Lee and Gyuho Lee and Joonhwan Lee},
title = {PromptPilot: Exploring User Experience of Prompting with AI-Enhanced Initiative in LLMs},
journal = {International Journal of Human–Computer Interaction},
pages = {1--23},
year = {2025},
publisher = {Taylor \& Francis},
doi = {10.1080/10447318.2025.2489030},
URL = {https://doi.org/10.1080/10447318.2025.2489030},
eprint = {https://doi.org/10.1080/10447318.2025.2489030}
}

@misc{lin2024promptopt,
      title={Prompt Optimization with Human Feedback}, 
      author={Xiaoqiang Lin and Zhongxiang Dai and Arun Verma and See-Kiong Ng and Patrick Jaillet and Bryan Kian Hsiang Low},
      year={2024},
      eprint={2405.17346},
      archivePrefix={arXiv},
      primaryClass={cs.LG},
      url={https://arxiv.org/abs/2405.17346}, 
}

@inproceedings{shen2025interactionaug,
  author    = {Shen, L. and Li, H. and Wang, Y. and Xie, X. and Qu, H.},
  title     = {Prompting Generative AI with Interaction-Augmented Instructions},
  booktitle = {Extended Abstracts of the CHI Conference on Human Factors in Computing Systems},
  year      = {2025},
  pages     = {1--9},
  publisher = {ACM},
  month     = apr
}

@misc{xiang2025selfsupervised,
      title={Self-Supervised Prompt Optimization}, 
      author={Jinyu Xiang and Jiayi Zhang and Zhaoyang Yu and Xinbing Liang and Fengwei Teng and Jinhao Tu and Fashen Ren and Xiangru Tang and Sirui Hong and Chenglin Wu and Yuyu Luo},
      year={2025},
      eprint={2502.06855},
      archivePrefix={arXiv},
      primaryClass={cs.CL},
      url={https://arxiv.org/abs/2502.06855}, 
}

@misc{wang2024langgptrethinkingstructuredreusable,
      title={LangGPT: Rethinking Structured Reusable Prompt Design Framework for LLMs from the Programming Language}, 
      author={Ming Wang and Yuanzhong Liu and Xiaoyu Liang and Songlian Li and Yijie Huang and Xiaoming Zhang and Sijia Shen and Chaofeng Guan and Daling Wang and Shi Feng and Huaiwen Zhang and Yifei Zhang and Minghui Zheng and Chi Zhang},
      year={2024},
      eprint={2402.16929},
      archivePrefix={arXiv},
      primaryClass={cs.SE},
      url={https://arxiv.org/abs/2402.16929}, 
}

@inproceedings{kim2023cellsgeneratorsandlenses,
author = {Kim, Tae Soo and Lee, Yoonjoo and Chang, Minsuk and Kim, Juho},
title = {Cells, Generators, and Lenses: Design Framework for Object-Oriented Interaction with Large Language Models},
year = {2023},
isbn = {9798400701320},
publisher = {Association for Computing Machinery},
address = {New York, NY, USA},
url = {https://doi.org/10.1145/3586183.3606833},
doi = {10.1145/3586183.3606833},
booktitle = {Proceedings of the 36th Annual ACM Symposium on User Interface Software and Technology},
articleno = {4},
numpages = {18},
location = {San Francisco, CA, USA},
series = {UIST '23}
}

@inproceedings{whyjohnnycantprompt,
author = {Zamfirescu-Pereira, J.D. and Wong, Richmond Y. and Hartmann, Bjoern and Yang, Qian},
title = {Why Johnny Can’t Prompt: How Non-AI Experts Try (and Fail) to Design LLM Prompts},
year = {2023},
isbn = {9781450394215},
publisher = {Association for Computing Machinery},
address = {New York, NY, USA},
url = {https://doi.org/10.1145/3544548.3581388},
doi = {10.1145/3544548.3581388},
booktitle = {Proceedings of the 2023 CHI Conference on Human Factors in Computing Systems},
articleno = {437},
numpages = {21},
location = {Hamburg, Germany},
series = {CHI '23}
}

@misc{mishra2025promptaidpromptexplorationperturbation,
      title={PromptAid: Prompt Exploration, Perturbation, Testing and Iteration using Visual Analytics for Large Language Models}, 
      author={Aditi Mishra and Utkarsh Soni and Anjana Arunkumar and Jinbin Huang and Bum Chul Kwon and Chris Bryan},
      year={2025},
      eprint={2304.01964},
      archivePrefix={arXiv},
      primaryClass={cs.HC},
      url={https://arxiv.org/abs/2304.01964}, 
}

@inproceedings{chainforge,
author = {Arawjo, Ian and Swoopes, Chelse and Vaithilingam, Priyan and Wattenberg, Martin and Glassman, Elena L.},
title = {ChainForge: A Visual Toolkit for Prompt Engineering and LLM Hypothesis Testing},
year = {2024},
isbn = {9798400703300},
publisher = {Association for Computing Machinery},
address = {New York, NY, USA},
url = {https://doi.org/10.1145/3613904.3642016},
doi = {10.1145/3613904.3642016},
booktitle = {Proceedings of the 2024 CHI Conference on Human Factors in Computing Systems},
articleno = {304},
numpages = {18},
location = {Honolulu, HI, USA},
series = {CHI '24}
}

@inproceedings{x3dobject,
author = {Polys, Nicholas and Mohammed, Ayat and Sandbrook, Ben},
title = {Prompt Engineering for X3D Object Creation with LLMs},
year = {2024},
isbn = {9798400706899},
publisher = {Association for Computing Machinery},
address = {New York, NY, USA},
url = {https://doi.org/10.1145/3665318.3677159},
doi = {10.1145/3665318.3677159},
booktitle = {Proceedings of the 29th International ACM Conference on 3D Web Technology},
articleno = {18},
numpages = {7},
location = {Guimar\~{a}es, Portugal},
series = {Web3D '24}
}

@misc{schulhoff2025promptreportsystematicsurvey,
      title={The Prompt Report: A Systematic Survey of Prompt Engineering Techniques}, 
      author={Sander Schulhoff and Michael Ilie and Nishant Balepur and Konstantine Kahadze and Amanda Liu and Chenglei Si and Yinheng Li and Aayush Gupta and HyoJung Han and Sevien Schulhoff and Pranav Sandeep Dulepet and Saurav Vidyadhara and Dayeon Ki and Sweta Agrawal and Chau Pham and Gerson Kroiz and Feileen Li and Hudson Tao and Ashay Srivastava and Hevander Da Costa and Saloni Gupta and Megan L. Rogers and Inna Goncearenco and Giuseppe Sarli and Igor Galynker and Denis Peskoff and Marine Carpuat and Jules White and Shyamal Anadkat and Alexander Hoyle and Philip Resnik},
      year={2025},
      eprint={2406.06608},
      archivePrefix={arXiv},
      primaryClass={cs.CL},
      url={https://arxiv.org/abs/2406.06608}, 
}

@inproceedings{oodrawing,
author = {Xia, Haijun and Araujo, Bruno and Grossman, Tovi and Wigdor, Daniel},
title = {Object-Oriented Drawing},
year = {2016},
isbn = {9781450333627},
publisher = {Association for Computing Machinery},
address = {New York, NY, USA},
url = {https://doi.org/10.1145/2858036.2858075},
doi = {10.1145/2858036.2858075},
booktitle = {Proceedings of the 2016 CHI Conference on Human Factors in Computing Systems},
pages = {4610–4621},
numpages = {12},
location = {San Jose, California, USA},
series = {CHI '16}
}

@misc{khot2023decomposedpromptingmodularapproach,
      title={Decomposed Prompting: A Modular Approach for Solving Complex Tasks}, 
      author={Tushar Khot and Harsh Trivedi and Matthew Finlayson and Yao Fu and Kyle Richardson and Peter Clark and Ashish Sabharwal},
      year={2023},
      eprint={2210.02406},
      archivePrefix={arXiv},
      primaryClass={cs.CL},
      url={https://arxiv.org/abs/2210.02406}, 
}

@misc{wei2023chainofthoughtpromptingelicitsreasoning,
      title={Chain-of-Thought Prompting Elicits Reasoning in Large Language Models}, 
      author={Jason Wei and Xuezhi Wang and Dale Schuurmans and Maarten Bosma and Brian Ichter and Fei Xia and Ed Chi and Quoc Le and Denny Zhou},
      year={2023},
      eprint={2201.11903},
      archivePrefix={arXiv},
      primaryClass={cs.CL},
      url={https://arxiv.org/abs/2201.11903}, 
}

@inproceedings{chainingllmvisual,
author = {Wu, Tongshuang and Jiang, Ellen and Donsbach, Aaron and Gray, Jeff and Molina, Alejandra and Terry, Michael and Cai, Carrie J},
title = {PromptChainer: Chaining Large Language Model Prompts through Visual Programming},
year = {2022},
isbn = {9781450391566},
publisher = {Association for Computing Machinery},
address = {New York, NY, USA},
url = {https://doi.org/10.1145/3491101.3519729},
doi = {10.1145/3491101.3519729},
booktitle = {Extended Abstracts of the 2022 CHI Conference on Human Factors in Computing Systems},
articleno = {359},
numpages = {10},
location = {New Orleans, LA, USA},
series = {CHI EA '22}
}

@inproceedings{aichains,
author = {Wu, Tongshuang and Terry, Michael and Cai, Carrie Jun},
title = {AI Chains: Transparent and Controllable Human-AI Interaction by Chaining Large Language Model Prompts},
year = {2022},
isbn = {9781450391573},
publisher = {Association for Computing Machinery},
address = {New York, NY, USA},
url = {https://doi.org/10.1145/3491102.3517582},
doi = {10.1145/3491102.3517582},
booktitle = {Proceedings of the 2022 CHI Conference on Human Factors in Computing Systems},
articleno = {385},
numpages = {22},
location = {New Orleans, LA, USA},
series = {CHI '22}
}

@inproceedings{reynolds2021prompt,
  title={Prompt programming for large language models: Beyond the few-shot paradigm},
  author={Reynolds, Laria and McDonell, Kyle},
  booktitle={Extended abstracts of the 2021 CHI conference on human factors in computing systems},
  pages={1--7},
  year={2021}
}

@misc{joshi2024coprompter,
      title={CoPrompter: User-Centric Evaluation of LLM Instruction Alignment for Improved Prompt Engineering}, 
      author={Ishika Joshi and Simra Shahid and Shreeya Venneti and Manushree Vasu and Yantao Zheng and Yunyao Li and Balaji Krishnamurthy and Gromit Yeuk-Yin Chan},
      year={2024},
      eprint={2411.06099},
      archivePrefix={arXiv},
      primaryClass={cs.HC},
      url={https://arxiv.org/abs/2411.06099}, 
}

@misc{cai2024lowcodellmgraphicaluser,
      title={Low-code LLM: Graphical User Interface over Large Language Models}, 
      author={Yuzhe Cai and Shaoguang Mao and Wenshan Wu and Zehua Wang and Yaobo Liang and Tao Ge and Chenfei Wu and Wang You and Ting Song and Yan Xia and Jonathan Tien and Nan Duan and Furu Wei},
      year={2024},
      eprint={2304.08103},
      archivePrefix={arXiv},
      primaryClass={cs.CL},
      url={https://arxiv.org/abs/2304.08103}, 
}

@misc{zhou2025instructpipegeneratingvisualblocks,
      title={InstructPipe: Generating Visual Blocks Pipelines with Human Instructions and LLMs}, 
      author={Zhongyi Zhou and Jing Jin and Vrushank Phadnis and Xiuxiu Yuan and Jun Jiang and Xun Qian and Kristen Wright and Mark Sherwood and Jason Mayes and Jingtao Zhou and Yiyi Huang and Zheng Xu and Yinda Zhang and Johnny Lee and Alex Olwal and David Kim and Ram Iyengar and Na Li and Ruofei Du},
      year={2025},
      eprint={2312.09672},
      archivePrefix={arXiv},
      primaryClass={cs.HC},
      doi={https://doi.org/10.1145/3706598.3713905},
      url={https://arxiv.org/abs/2312.09672}, 
}

@misc{li2025ipropinteractivepromptoptimization,
      title={iPrOp: Interactive Prompt Optimization for Large Language Models with a Human in the Loop}, 
      author={Jiahui Li and Roman Klinger},
      year={2025},
      eprint={2412.12644},
      archivePrefix={arXiv},
      primaryClass={cs.CL},
      url={https://arxiv.org/abs/2412.12644}, 
}

@inproceedings{graphologue,
author = {Jiang, Peiling and Rayan, Jude and Dow, Steven P. and Xia, Haijun},
title = {Graphologue: Exploring Large Language Model Responses with Interactive Diagrams},
year = {2023},
isbn = {9798400701320},
publisher = {Association for Computing Machinery},
address = {New York, NY, USA},
url = {https://doi.org/10.1145/3586183.3606737},
doi = {10.1145/3586183.3606737},
booktitle = {Proceedings of the 36th Annual ACM Symposium on User Interface Software and Technology},
articleno = {3},
numpages = {20},
location = {San Francisco, CA, USA},
series = {UIST '23}
}

@misc{mao2025promptstemplatessystematicprompt,
      title={From Prompts to Templates: A Systematic Prompt Template Analysis for Real-world LLMapps}, 
      author={Yuetian Mao and Junjie He and Chunyang Chen},
      year={2025},
      eprint={2504.02052},
      archivePrefix={arXiv},
      primaryClass={cs.SE},
      url={https://arxiv.org/abs/2504.02052}, 
}

@ARTICLE{directmanipulation,
  author={Shneiderman},
  journal={Computer}, 
  title={Direct Manipulation: A Step Beyond Programming Languages}, 
  year={1983},
  volume={16},
  number={8},
  pages={57-69},
  doi={10.1109/MC.1983.1654471}}

@inproceedings{sensecape, series={UIST ’23},
   title={Sensecape: Enabling Multilevel Exploration and Sensemaking with Large Language Models},
   url={http://dx.doi.org/10.1145/3586183.3606756},
   DOI={10.1145/3586183.3606756},
   booktitle={Proceedings of the 36th Annual ACM Symposium on User Interface Software and Technology},
   publisher={ACM},
   author={Suh, Sangho and Min, Bryan and Palani, Srishti and Xia, Haijun},
   year={2023},
   month=oct, pages={1–18},
   collection={UIST ’23} }

@inproceedings{10.1145/263552.263612,
author = {Mackay, Wendy E. and Fayard, Anne-Laure},
title = {HCI, natural science and design: a framework for triangulation across disciplines},
year = {1997},
isbn = {0897918630},
publisher = {Association for Computing Machinery},
address = {New York, NY, USA},
url = {https://doi.org/10.1145/263552.263612},
doi = {10.1145/263552.263612},
booktitle = {Proceedings of the 2nd Conference on Designing Interactive Systems: Processes, Practices, Methods, and Techniques},
pages = {223–234},
numpages = {12},
location = {Amsterdam, The Netherlands},
series = {DIS '97}
}

@article{thematicanalysis,
  title={Thematic analysis: A practical guide},
  author={Braun, Virginia and Clarke, Victoria},
  year={2021},
  publisher={SAGE publications Ltd}
}

@misc{ma2023fairnessguidedfewshotpromptinglarge,
      title={Fairness-guided Few-shot Prompting for Large Language Models}, 
      author={Huan Ma and Changqing Zhang and Yatao Bian and Lemao Liu and Zhirui Zhang and Peilin Zhao and Shu Zhang and Huazhu Fu and Qinghua Hu and Bingzhe Wu},
      year={2023},
      eprint={2303.13217},
      archivePrefix={arXiv},
      primaryClass={cs.CL},
      url={https://arxiv.org/abs/2303.13217}, 
}

@book{booch2007oop,
  title={Object-Oriented Analysis and Design with Applications},
  author={Booch, Grady and Maksimchuk, Robert and Engle, Michael and Young, Bobbi and Conallen, Jim and Houston, Kelli},
  year={2007},
  publisher={Addison-Wesley Professional},
  edition={3rd}
}

@book{the_design_of_everyday_things,
  title={The design of everyday things: Revised and expanded edition},
  author={Norman, Don},
  year={2013},
  publisher={Basic books}
}

@misc{wu2023autogenenablingnextgenllm,
      title={AutoGen: Enabling Next-Gen LLM Applications via Multi-Agent Conversation}, 
      author={Qingyun Wu and Gagan Bansal and Jieyu Zhang and Yiran Wu and Beibin Li and Erkang Zhu and Li Jiang and Xiaoyun Zhang and Shaokun Zhang and Jiale Liu and Ahmed Hassan Awadallah and Ryen W White and Doug Burger and Chi Wang},
      year={2023},
      eprint={2308.08155},
      archivePrefix={arXiv},
      primaryClass={cs.AI},
      url={https://arxiv.org/abs/2308.08155}, 
}

@inproceedings{hutchinson2003technology,
  title={Technology probes: inspiring design for and with families},
  author={Hutchinson, Hilary and Mackay, Wendy and Westerlund, Bo and Bederson, Benjamin B and Druin, Allison and Plaisant, Catherine and Beaudouin-Lafon, Michel and Conversy, St{\'e}phane and Evans, Helen and Hansen, Heiko and others},
  booktitle={Proceedings of the SIGCHI conference on Human factors in computing systems},
  pages={17--24},
  year={2003}
}

@misc{ramesh2022,
      title={Hierarchical Text-Conditional Image Generation with CLIP Latents}, 
      author={Aditya Ramesh and Prafulla Dhariwal and Alex Nichol and Casey Chu and Mark Chen},
      year={2022},
      eprint={2204.06125},
      archivePrefix={arXiv},
      primaryClass={cs.CV},
      url={https://arxiv.org/abs/2204.06125}, 
}

@article{betker2023,
  title={Improving image generation with better captions},
  author={Betker, James and Goh, Gabriel and Jing, Li and Brooks, Tim and Wang, Jianfeng and Li, Linjie and Ouyang, Long and Zhuang, Juntang and Lee, Joyce and Guo, Yufei and others},
  journal={Computer Science. https://cdn. openai. com/papers/dall-e-3. pdf},
  volume={2},
  number={3},
  pages={8},
  year={2023}
}

@inproceedings{zhang2023adding,
  title={Adding conditional control to text-to-image diffusion models},
  author={Zhang, Lvmin and Rao, Anyi and Agrawala, Maneesh},
  booktitle={Proceedings of the IEEE/CVF international conference on computer vision},
  pages={3836--3847},
  year={2023}
}

@inproceedings{han2025poet,
  title={POET: Supporting Prompting Creativity and Personalization with Automated Expansion of Text-to-Image Generation},
  author={Han, Evans Xu and Zhang, Alice Qian and Zhu, Haiyi and Shen, Hong and Liang, Paul Pu and Hsieh, Jane},
  booktitle={Proceedings of the 38th Annual ACM Symposium on User Interface Software and Technology},
  pages={1--18},
  year={2025}
}

@inproceedings{masson2024directgpt,
  title={Directgpt: A direct manipulation interface to interact with large language models},
  author={Masson, Damien and Malacria, Sylvain and Casiez, G{\'e}ry and Vogel, Daniel},
  booktitle={Proceedings of the 2024 CHI Conference on Human Factors in Computing Systems},
  pages={1--16},
  year={2024}
}

@inproceedings{xu2024jamplate,
  title={Jamplate: Exploring llm-enhanced templates for idea reflection},
  author={Xu, Xiaotong and Yin, Jiayu and Gu, Catherine and Mar, Jenny and Zhang, Sydney and E, Jane L and Dow, Steven P},
  booktitle={Proceedings of the 29th International Conference on Intelligent User Interfaces},
  pages={907--921},
  year={2024}
}

@article{jiang2020can,
  title={How can we know what language models know?},
  author={Jiang, Zhengbao and Xu, Frank F and Araki, Jun and Neubig, Graham},
  journal={Transactions of the Association for Computational Linguistics},
  volume={8},
  pages={423--438},
  year={2020},
  publisher={MIT Press One Rogers Street, Cambridge, MA 02142-1209, USA journals-info~…}
}

@inproceedings{mishra2022reframing,
  title={Reframing instructional prompts to gptk’s language},
  author={Mishra, Swaroop and Khashabi, Daniel and Baral, Chitta and Choi, Yejin and Hajishirzi, Hannaneh},
  booktitle={Findings of the association for computational linguistics: ACL 2022},
  pages={589--612},
  year={2022}
}

@inproceedings{subramonyam2024bridging,
  title={Bridging the gulf of envisioning: Cognitive challenges in prompt based interactions with llms},
  author={Subramonyam, Hari and Pea, Roy and Pondoc, Christopher and Agrawala, Maneesh and Seifert, Colleen},
  booktitle={Proceedings of the 2024 CHI Conference on Human Factors in Computing Systems},
  pages={1--19},
  year={2024}
}

@inproceedings{petridis2023promptinfuser,
  title={Promptinfuser: Bringing user interface mock-ups to life with large language models},
  author={Petridis, Savvas and Terry, Michael and Cai, Carrie Jun},
  booktitle={Extended Abstracts of the 2023 CHI Conference on Human Factors in Computing Systems},
  pages={1--6},
  year={2023}
}

@inproceedings{kim2024evallm,
  title={Evallm: Interactive evaluation of large language model prompts on user-defined criteria},
  author={Kim, Tae Soo and Lee, Yoonjoo and Shin, Jamin and Kim, Young-Ho and Kim, Juho},
  booktitle={Proceedings of the 2024 CHI Conference on Human Factors in Computing Systems},
  pages={1--21},
  year={2024}
}

@misc{ma2024you,
  title={What you say= what you want? Teaching humans to articulate requirements for LLMs},
  author={Ma, Qianou and Peng, Weirui and Shen, Hua and Koedinger, Kenneth and Wu, Tongshuang},
  year={2024}
}

@article{ma2025should,
  title={What should we engineer in prompts? training humans in requirement-driven llm use},
  author={Ma, Qianou and Peng, Weirui and Yang, Chenyang and Shen, Hua and Koedinger, Ken and Wu, Tongshuang},
  journal={ACM Transactions on Computer-Human Interaction},
  volume={32},
  number={4},
  pages={1--27},
  year={2025},
  publisher={ACM New York, NY}
}

@inproceedings{chen2025genui,
  title={The GenUI Study: Exploring the Design of Generative UI Tools to Support UX Practitioners and Beyond},
  author={Chen, Xiang'Anthony and Knearem, Tiffany and Li, Yang},
  booktitle={Proceedings of the 2025 ACM Designing Interactive Systems Conference},
  pages={1179--1196},
  year={2025}
}

@article{sarkar2025conversational,
  title={Conversational user-ai intervention: A study on prompt rewriting for improved llm response generation},
  author={Sarkar, Rupak and Sarrafzadeh, Bahareh and Chandrasekaran, Nirupama and Rangan, Nagu and Resnik, Philip and Yang, Longqi and Jauhar, Sujay Kumar},
  journal={arXiv preprint arXiv:2503.16789},
  year={2025}
}

@inproceedings{liu2025user,
  title={User feedback in human-LLM dialogues: a lens to understand users but noisy as a learning signal},
  author={Liu, Yuhan and Zhang, Michael JQ and Choi, Eunsol},
  booktitle={Proceedings of the 2025 Conference on Empirical Methods in Natural Language Processing},
  pages={2666--2681},
  year={2025}
}

@article{achiam2023gpt,
  title={Gpt-4 technical report},
  author={Achiam, Josh and Adler, Steven and Agarwal, Sandhini and Ahmad, Lama and Akkaya, Ilge and Aleman, Florencia Leoni and Almeida, Diogo and Altenschmidt, Janko and Altman, Sam and Anadkat, Shyamal and others},
  journal={arXiv preprint arXiv:2303.08774},
  year={2023}
}

@article{team2023gemini,
  title={Gemini: a family of highly capable multimodal models},
  author={Team, Gemini and Anil, Rohan and Borgeaud, Sebastian and Alayrac, Jean-Baptiste and Yu, Jiahui and Soricut, Radu and Schalkwyk, Johan and Dai, Andrew M and Hauth, Anja and Millican, Katie and others},
  journal={arXiv preprint arXiv:2312.11805},
  year={2023}
}
